%% file: paper.tex
\RequirePackage{fix-cm}

\documentclass[smallextended]{svjour3}  
\usepackage{subcaption}
\usepackage{placeins}

\smartqed  % flush right qed marks, e.g. at end of proof
\usepackage[numbers]{natbib}
\usepackage{graphicx}
\usepackage[linesnumbered,ruled]{algorithm2e}

\usepackage{amsmath}

\usepackage{mathptmx}      % use Times fonts if available on your TeX system
\usepackage{url}

\usepackage[usenames,dvipsnames]{xcolor}

\usepackage{tikz}

\usetikzlibrary{decorations.pathreplacing}

\usepackage{fancyhdr}

\begin{document}
%\fancyhf{}
\pagestyle{fancy}
\fancyfoot[C]{\footnotesize Preprint of Chuan C.H., Agres K., Herremans D. In Press. From Context to Concept: Exploring Semantic Relationships in Music with Word2Vec. \emph{Neural Computing and Applications}, Springer. }
\title{From Context to Concept: Exploring Semantic Relationships in Music with Word2Vec}
%From Context to Concept: Exploring The Semantic Meaning of Music \kat{in?} Word2Vec Space... \kat{Or how about: From Context to Concept: Exploring Semantic Relationships in Music with Word2Vec}}
%\title{Modeling Musical Context Using Word2vec}

%\thanks{Preprint of Chuan C.H., Agres K., Herremans D. In Press (2018). Neural Computing and Applications, Springer.}

%Grants or other notes
%\subtitle{subtitle}


\author{Ching-Hua Chuan         \and
        Kat Agres \and Dorien Herremans %etc.
}


\institute{Ching-Hua Chuan \at
              Department of Cinema \& Interactive Media, School of Communication, University of Miami, USA \\ 
              Tel.: +1-305-2844388\\
              Fax: +1-305-2845226\\
              \email{c.chuan@miami.edu}           %  \\
           \and
           Kat Agres\at Social and Cognitive Computing Department, Institute for High Performance Computing, A*STAR, Singapore
%\kat{I'm listed last here, but second above. I don't mind which order, but we should be consistent. Dorien, would you like to be last author or second?} \ch{You two can decide. I am fine either way.} \dorien{ok like this?} 
\and
           Dorien Herremans\at
              Information Systems, Technology, and Design Department, Singapore University of Technology and Design, Singapore \\ Social and Cognitive Computing Department, Institute for High Performance Computing, A*STAR, Singapore           
}

\date{Received: 22/06/2018 / Accepted: 26/11/2018}

\maketitle

\begin{abstract}

%\rewrite{OLD: We present a semantic vector space model for capturing complex polyphonic musical context. A word2vec model based on a skip-gram representation with negative sampling was used to model slices of music from a dataset of Beethoven's piano sonatas. A visualization of the reduced vector space using t-distributed stochastic neighbor embedding shows that the resulting embedded vector space captures tonal relationships, even without any explicit information about the musical contents of the slices. Secondly, an excerpt of the Moonlight Sonata from Beethoven was altered by replacing slices based on context similarity. The resulting music shows that the selected slice based on similar word2vec context also has a relatively short tonal distance from the original slice.} 
%kat: 
We explore the potential of a popular distributional semantics vector space model, \textit{word2vec}, for capturing meaningful relationships in ecological (complex polyphonic) music. More precisely, the skip-gram version of word2vec is used to model slices of music from a large corpus spanning eight musical genres. In this newly learned vector space, a metric based on cosine distance is able to distinguish between functional chord relationships, as well as harmonic associations in the music. Evidence, based on cosine distance between chord-pair vectors, suggests that an implicit circle-of-fifths exists in the vector space. In addition, a comparison between pieces in different keys reveals that key relationships are represented in word2vec space. These results suggest that the newly learned embedded vector representation does in fact capture tonal and harmonic characteristics of music, without receiving explicit information about the musical content of the constituent slices. In order to investigate whether proximity in the discovered space of embeddings is indicative of `semantically-related' slices, we explore a music generation task, by automatically replacing existing slices from a given piece of music with new slices. We propose an algorithm to find substitute slices based on spatial proximity and the pitch class distribution inferred in the chosen subspace. The results indicate that the size of the subspace used has a significant effect on whether slices belonging to the same key are selected. 
In sum, the proposed word2vec model is able to learn music-vector embeddings that capture meaningful tonal and harmonic relationships in music, thereby providing a useful tool for exploring musical properties and comparisons across pieces, as a potential input representation for deep learning models, and as a music generation device.

%\kat{The last three sentences are all about generation/slice replacement, but I don't think that's the main thrust of the paper (and we don't have evaluation of this section). Maybe we should add another sentence about tonal relationships, or about analogy, before the sentence starting `Finally, to..'?} 
%\dorien{We don't test the algorithm on that many slices either so I'm not sure we should talk so much about it?} \dorien{Agreed that we could focus more on other results. Analogies is the circle of fifths though. I've added a little bit.} \kat{True, but I just meant not over-emphasizing the slice replacement aspect. I don't think the abstract ends strongly, so I'll add a concluding sentence.} \dorien{Great sentence. I like it, ok for me. } \ch{The ending of the abstract is good!}

%the musical context and \kat{[Ching-Hua, for your input, in case you have a specific wording in mind:} low cosine similarity etc] in a musical generation task.
%\kat{We should talk about genre classification too?}

%Insert your abstract here. Include keywords, PACS and mathematical
%subject classification numbers as needed.
% \subclass{MSC code1 \and MSC code2 \and more} music context, word2vec, music, neural networks, semantic vector space 
\end{abstract}

%\dorien{I think our reference section is still quite limited, wouldn't hurt to cite a bit more} \kat{Sure. I added a few references this weekend, but will keep a lookout for good places to add more.}
%\dorien{this is for genre classification, did not use: mention \url{https://radimrehurek.com/gensim/models/doc2vec.html}}
%\item choose new dataset
%\item run model again
%\item redo tsne graphs, with different musical properties/labeling?
%\item test classification? Which type? E.g. from a known classification dataset? MIREX?
%\item subspace model from Kat? and create subspaces based on chord progressions?
%\end{itemize}

\section{Introduction}
%\kat{Finished first stab at new introduction.}\dorien{cool thanks!}

Distributional semantic vector space models have become important modeling tools in computational linguistics and natural language processing (NLP)~\citep{liddy1999multilingual, turney2010frequency}. 
These models
%~\citep{rumelhart1988learning} 
are typically used to represent (or embed) words as vectors in a continuous, multi-dimensional vector space, in which geometrical relationships between word vectors are significant~\citep{agres2016modeling, liddy1999multilingual, mcgregor2015distributional, turney2010frequency}. For example, semantically similar words tend to occur in close proximity within the space, and analogical relationships may be discovered in some cases as vectors with similar angle and distance properties ~\citep{turney2010frequency}. 
A popular approach to creating vector spaces for NLP is a neural network-based approach called \textit{word2vec}~\citep{mikolov2013distributed}.
In this paper, we explore whether word2vec can be used to model a related form of auditory expression: music.
We build upon the previous work of the authors \citep{herremans2017word2vec}, %\cite{herremans2017word2vec}, \ch{if we submit to the special issue, I wonder if we should do it anonymously.} \dorien{This may be a good idea in fact... the paper with my name on it couldn't get reviewers! I think people realised the conflict} \kat{agreed. How about we put: ``[made anonymous for submission]'' where the reference should be?}\ch{Let's do this!}
in which a preliminary model was built by training word2vec on a small music dataset. This research takes a more comprehensive approach to exploring the extent to which word2vec is capable of discovering different types of meaningful features of music, such as tonal and harmonic relationships.

In the field of music cognition, there is a long tradition of investigating how the statistical properties of music influence listeners' perceptual and emotional responses. For example, the frequency of occurrence of tones in a key helps shape the perception of the tonal hierarchy (e.g., the relative stability of different notes in the key)~\cite{krumhansl1990cognitive}, and the likelihood that a particular tone or chord will follow a previous tone(s) or chord(s) helps drive tonal learning and expectation mechanisms~\cite{saffran1999statistical, pearce2012auditory, agres2018information}, as well as emotional responses to music~\cite{huron2006sweet, meyer1956emotion}.
Researchers have employed various methods to capture the statistical properties of music using machine learning techniques, including Markov models~\citep{conklin1995multiple}, Recursive Neural Networks (RNNs) combined with Restricted Bolzmann Machines~\citep{boulanger2012modeling}, and Long-Short Term RNN models~\citep{cancino2017bach, chuan2017lstm, eck2002finding, sak2014long}. 
Notably, these models all use input representations that contain explicit information about musical structure (e.g. pitch intervals, chords, keys, etc). 
%\ch{we also use pitch class in this paper. Can we say the previous work uses chords or keys?} \kat{Changed this.}
%In this research, rather than relying upon the explicit musical content, our goal is to use the \emph{context} to discover meaningful relationships in the music. 
%\kat{[I'm not sure I completely agree with this statement, it's the content of the context, right? How about instead]: 
In the present research, rather than relying upon this sort of explicit musical content (e.g., defining the meaning inherent in particular notes and chords, such as their relative tonal stability values), we use distributional semantics techniques to examine how \emph{contextual} information (e.g., the surrounding set of musical events%, as distinctive from the content and associated meaning of these events
) may be used to discover musical meaning. The premise of this approach is that examining the surrounding context and co-occurrence statistics of musical events can yield insight into the semantics (in this case, musical relationships) of the domain.

%\ch{Do we want talk about how important/useful/popular word2vec is in deep learning? Since this is a special issue on deep learning... Also, I think the advantage that deep learning provides is the so-called end-to-end solution, which means no need for hand-crafted features. If word2vec can pick up some musical meanings, it can be incorporated in the DL solution so that no need for manual labeling of musical elements. What do you think?} \dorien{We may mention this yeah, that vectors could serve as input for lstm or so, we may also want to write this in future research} \kat{Just cleaned this up a bit; thanks for adding.} \ch{It looks good. Thanks! Just added the citations.}
%\dorien{something like:} 
In NLP research employing deep learning, vector space models such as word2vec play an important role, because they provide a dense vector representation of words. In this research, we explore the use of word2vec in the context of music. Computational models of music often use convolutional neural networks (CNNs)\cite{chuan2017lstm, korzeniowski2016fully}, which are renowned for their ability to `learn' features automatically~\cite{kim2014convolutional, poria2015deep, poria2016deeper}. These CNNs are typically combined with other models, such as memory-based neural networks like LSTMs~\cite{chuan2017lstm} and RNNs~\cite{boulanger2012modeling}, thus providing an end-to-end solution without the need for hand-crafted features. Therefore, the word2vec representation proposed in this research may provide an alternative input for RNNs and LSTMs in the domain of music.

Few researchers have attempted to model musical context using semantic vector space models. \citet{huang2016chordripple} used word2vec to model chord progressions as a method of discovering harmonic relationships and
%\kat{(discovering harmonic relationships and?)}\dorien{ok I think} 
recommending chords to novice composers. \citet{madjiheurem2016chord2vec} also trained NLP-inspired models for chord sequences. In their preliminary work, only training results were compared, not the actual ability of the models to capture musical properties. 
%http://www.cs.nott.ac.uk/~psztg/cml/2016/papers/CML2016_paper_5.pdf.
 Using chords as input, however, gives the model explicit musical content, and vastly reduces the extensive space of all musical possibilities to a very small vocabulary. In this paper, complex polyphonic music is represented as a sequence of `slices' that have undergone no pre-processing to extract musical features, such as key signature or time signature. Our goal is to explore the extent to which word2vec is able to discover semantic similarity and musical relationships by looking only at the \textit{context} in which every musical slice appears. 
We might adapt the famous saying by linguist J. R. \citet{firth1957synopsis} from, ``You shall know a word by the company it keeps!'' to ``You shall know a slice of music by the company it keeps'', where a slice of music is a note or chord within a particular time interval. %\ch{I like the sentence! But we also need to make sure that slices in our study do not necessary mean chords.} \dorien{something like that?} \kat{From discussions at QM, I'm slightly wary to draw false equivalence between words and chords.. Have slightly updated this :) I realize `musical event' doesn't have the same ring to it as chord, but what do you think?} \dorien{ok for me!} \ch{How about `You shall know a note or chord by the company it keeps'? } \kat{What about the original idea of 'musical slice'? So how about this: ``You shall know a musical slice by the company it keeps'', where a slice of music is a note or chord within a particular time interval.} \dorien{ok for me!} \ch{sounds good!}

The structure of the paper will be as follows: the next sections describe the implementation of the word2vec model, followed by the way in which the music is represented in this study. We then report on the model's ability to capture tonal relationships, and then %\ch{Currently I put the section on music generation the last. Shall we change the words here to reflect the order, or do you want to move the section to change the order?} DH: ok changed. 
empirically test the model's ability to capture key and harmonic relationships (including analogical relationships across keys). The results section discusses the use of the model for music generation, and is followed by the conclusion. %\ch{I know we talked about putting music generation after the section about chord, but I decided to move it to the last section because generation is very different from the other analyses (chord/key/analogy) procedurally. What do you think?} \kat{Yeah, I see what you mean. This organization also works.}

% \dorien{Useful quote? linguist J. R. Firth (1957), who phrased it as “You shall know a word by the company it keeps!”.}
% \kat{Yeah, this is the most famous "definition" of the dist semantics hypothesis. We can include it in the Intro if you like.}

%\dorien{are there others in music? I recently reviewed a paper for ICCC but it's not published yet}
%\ch{I think word2vec was introduced in NLP mainly to eliminate the curse of dimensionality. Using word2vec is helpful when modeling polyphonic musical slices because the pure bag-of-word representation for musical slices will result in too many unique 'words' or slices, which becomes the problem of sparse vector representation. The situation is worse when music is sliced in micro-beat level.}

%If it fits somewhere: NLP model based on word2vec (but different) \citep{mcgregor2015distributional}

%"NLP, but all methods depend in some way or another on the Distributional Hypothesis, which states that words that appear in the same contexts share semantic meaning. \ch{Do you want to add more about NLP? Or maybe a transition to introduce the following sections?} 
%Predictive models directly try to predict a word from its neighbors in terms of learned small, dense embedding vectors (considered parameters of the model).

%Testing if this context captures the tonal similarity (it does not). 

%Show a cloud  and music with substitutions

%\dorien{I've been looking up the capitalization of word2vec, I don't think we need to capitalise it? Or maybe just the first letter... } \ch{probably not capitalize it at all}

\section{Word2vec}

The SMART document information retrieval system by~\citet{salton1971smart} and~\citet{salton1975vector} was the first to utilize a vector space model. In this initial work, each document in a collection was represented as a dot in a multidimensional vector space. A user query was also represented as a dot in this same space. The query results included the closest documents to the query word in the vector space, as they were assumed to have similar meaning. This initial model was developed through building a frequency matrix. 
Traditional methods for creating vector space models typically take a bag-of-words approach~\cite{schwartz2015symmetric} and create a vector representation for each word $w_i$ based on the co-occurrence frequency with each other word $w_j$. This initial representation can then be post-processed using methods such as Positive Pointwise Mutual Information~\cite{erk2012vector,mcgregor2015distributional}, normalization, and dimensionality reduction~\cite{dhillon2011multi, lebret2013word}. Another popular model is called GloVe~\cite{pennington2014glove}, short for Global Vectors. This model also starts from a co-occurrence matrix and then trains a log-bilinear regression model to generate embeddings on the non-zero elements. 
Subsequently, in the early 2000s, an interest emerged in using neural network techniques such as word2vec for building vector space models~\cite{bengio2003neural, collobert2008unified, collobert2011natural, mikolov2013linguistic, mnih2009scalable, mnih2013learning}.

%\dorien{the above better now?}
%\kat{Dorien, was this thought in relation to neural nets vs co-occurrence stats? or skip-gram vs cbow?} %\dorien{neural nets vs co-occurrence stats, or basically, just any history of how to make vector space models}
%to developing semantic vector space models. One method uses co-occurrence statistics of words to build the vector space [ref], while as the second method relies on neural networks [ref]. In this paper, we focus on the latter approach and implement a word2vec model for building our vector space \dorien{Why do we use this methods, is this better then co-occurrence? any reason? or I can just say it's popular and effective. } \kat{But that's the thing, the neural network is also effectively learning co-occurrence statistics, it's just using a different method to do so..} \dorien{ok, so I should rewrite to focus on the co-occurence technique. Will need to look up some history then} \kat{I'm confused by what you mean by "co-occurrence technique".. A neural net is not computing co-occurrence statistics in the same way, of course, because it is only selecting one of the context words (and it's a neural net, it's not counting co-occurrences or PMI or similar).} \dorien{I'll rewrite the above. We need some history on which type of models have been used. I see you have some in your paper Kat}

In this research, we work with  word2vec~\cite{mikolov2013distributed}, which provides an efficient way to create semantic vector spaces. A simple neural network model is trained on a corpus of text, in order to quickly create a vector space that can easily consist of several hundred dimensions~\citep{mikolov2013efficient}. In this multi-dimensional space, each word that occurs in the corpus can be represented as a vector. 
This semantic vector space reflects the distributional semantics hypothesis, which states that words that appear in the same contexts tend to have similar meanings \cite{harris1954distributional}. In terms of vector spaces, this means that words that occur in the same contexts appear geometrically close to one another. This allows us to assess semantic similarity of words based on geometric distance metrics such as cosine distance.

\paragraph{Skip-gram with negative sampling} 

\begin{figure}[h]
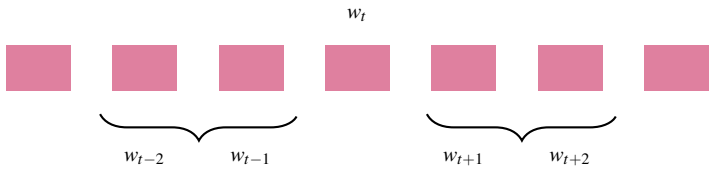
 \centering
\include{graph}
\caption{Example of how a skip gram window of size $c = 4$ is determined on word $w_t$ given a sequence of words.}
\label{fig:skipgram}
\end{figure}

Two types of approaches are typically used when building word2vec models: continuous bag-of-words (CBOW) and skip-gram. With a CBOW approach, the surrounding words are used to predict the current word~\citep{mikolov2013exploiting}, while the skip gram-model takes a current word and tries to predict the words in a surrounding window of size $c$ (see Figure~\ref{fig:skipgram}). 

Both models are computationally efficient and have low complexity, which means that they can both handle a corpus of billions of words without much effort. \citet{mikolov2013efficient} states that while CBOW models are often slightly faster, skip-gram models tend to perform better on smaller datasets. In this research, we used the latter approach for the dataset described in Section ~\ref{sec:dataset}.%In this research, we therefore used the latter approach. 

In the current implementation, the neural network takes one word $w_t$ as input and tries to predict one of the surrounding words $w_{t+i}$. The input-output pairs are randomly sampled $k$-times by selecting a random index value $i$ from the full skip-gram window $c$. All of the sampled values for $i$ for one input word $w_t$ form the set $J$. We can thus define the training objective as: 

\begin{equation}
\frac{1}{T}\sum_{t=1}^T \sum_{\forall i \in J} \log p(w_{t+i} | w_t),
\end{equation}

%\ch{For this equation, can we change the second summation from for all i in j to i = -c/2 to c/2? This way we don't have to use J.} \dorien{I changed it. I removed the forall too, that's still correct I think?}\ch{Yes, it is.}  \dorien{Actually I just realised we should use the old formula, because from the window, the same value could be sampled more then once, and not all values in the window are necessarily included. } \ch{got it!}

A softmax function is used to calculate term $p(w_{t+i} | w_t)$, however, it is computationally expensive to calculate the gradient of this term. Techniques that allow  this problem to be avoided include noise contrastive estimation~\citep{gutmann2012noise} and negative sampling~\cite{mikolov2013distributed}. In this research we implement the latter.

Negative sampling builds upon the idea that a trained model should be able to differentiate data from noise~\citep{goldberg2014word2vec}. Based on this assumption, the training objective is approximated by formulating a more efficient implementation in which a binary logistic regression is used to classify real data versus noise samples. The objective is to maximize the assigned probability of real words and minimize that of noise samples~\cite{mikolov2013distributed}.

%https://arxiv.org/pdf/1402.3722.pdf

%\dorien{cosine distance, describe also, indicate which we used. }
%\ch{I only use cosine distance now. So I changed the formula from cosine similarity to cosine distance.}

\paragraph{Cosine distance} D$_s$(A, B) is used to determine the distance between two words, represented by non-zero vectors A and B, in an $n$-dimensional vector space. The angle between A and B is defined as $\theta$. The cosine similarity metric can be calculated as follows: 

\begin{equation}
\text{D$_c$(A, B)} = 1- cos(\theta) = 1- \text{D$_s$(A, B)}
\end{equation}
where D$_s$(A, B) refers to \emph{cosine similarity}, a term often used to indicate the complement of cosine distance~\citep{Tan:2005}:

%between two non-zero vectors of an inner product space that measures the cosine of the angle between them

\begin{equation}
\text{D$_s$(A, B)} = \frac{\sum_{i=1}^n{A_i \times B_i}}{\sqrt{\sum_{i=1}^n{A^2_i}} \times \sqrt{\sum_{i=1}^n{B^2_i}}}
\end{equation}

% Cosine distance is a term often used for the complement in positive space, that is: D C ( A , B ) = 1 − S C ( A , B ) , {\displaystyle D_{C}(A,B)=1-S_{C}(A,B),} {\displaystyle D_{C}(A,B)=1-S_{C}(A,B),} where D C {\displaystyle D_{C}} D_C is the cosine distance and S C {\displaystyle S_{C}} S_{C} is the cosine similarity.

The following section describes the representation that we used to port the word2vec model to the domain of music.

\section{Musical slices as words}
\label{sec:slice}

%\dorien{We can talk a bit about the similarities of music and language, \kat{on it.} as per \cite{besson2001comparison}:
%\dorien{A last point that we could add, but we dont' ahve to... :-Common ancestor of music and language: expression of emotive meaning
%    e.g. emotional excitement: fast, accelerating, and high-pitched sound patterns}
%-Sequential elements that unfold in time: Rhythm
%- Temporal ratios (notes/phonemes, chords/words)
%- Grammar: musical grammar can include contour, cadence for closing, etc (more flexible)
%- Ability to generate strong expectancies: both unexpected (incongruent) words and notes generate peaks measurable in brain potentials (N400 \& P600)

%Although music lacks the explicit referential semantics of language, the two domains possess certain similar characteristics; for example, both contain structural aspects that can be thought of as a grammar or set of rules that guide the perceiver's expectations \cite{besson2001comparison}.
Although music lacks the precise referential semantics of language, there are clear similarities between these two expressive domains. Like language, music contains sequential events that are combined or constrained according to what can be thought of as a set of grammatical rules (although these rules are, of course, often more flexible for music than language). %guide the perceiver's expectations \cite{besson2001comparison}.
In addition, both domains are structured in such a way that they generate \textit{expectations} about forthcoming events (words or chords, etc). Unexpected words or notes have both been shown to elicit ``unexpectedness'' responses in the brain, such as the N400 or early right anterior negativity (ERAN) event-related potentials (ERPs) \cite{besson2001comparison,koelsch2002effects}. 
%\kat{How is this so far? About what you were looking for? I can add more..} \ch{This is great! Maybe we need another paragraph talking about notes/chords in music versus words in language and the difference between them. This will create a nice transition to why we process music into slices.} \kat{Okay, I've started on this below..}
In language, grammatical rules place constraints upon \textit{word} order. In music, the style and music-theoretic rules influence the order of \textit{musical events}, such as tones or chords. Unlike language, of course, multiple musical elements can be sounded simultaneously, and may even convey similar semantics (for example, instead of a C major chord containing the pitches of C, E and G, a high C one octave above the lower C may be added, without altering the chord's classification as C major). Because of this feature of music, in which multiple events may be presented simultaneously in addition to sequentially, we propose for the purpose of this research that the smallest unit of naturalistic music is the `\textit{musical slice}' - a brief duration (contingent on the time signature, musical density, etc) in which one or more tones may occur.
%(e.g., not using a score-based representation)
%\kat{CH, is this sort of what you were looking for? I haven't added references - I think DH had some in mind?} \dorien{Great! I was mostly thinking of the paper \cite{besson2001comparison}, which discusses all of my points above. }\ch{Looks great!}

In order to investigate the extent to which word2vec is able to model musical context, and how much \emph{semantic meaning in music} can be learned from the context, polyphonic music is represented with as little added musical knowledge as possible. We segment each piece in the corpus into equal-length slices, each consisting of a list of pitch classes contained within that slice. More specifically, each piece is segmented into beat-long, non-overlapping slices (note that different pieces can have different beat durations).
%\kat{Sorry, just noticed this - I thought the slices were equal length within the piece, but not necessarily across different pieces? Or did we change that?} \ch{You are right. Each piece is segmented into beat-long slices, but different pieces can have different lengths for beats.} \kat{Updated the text to reflect this point more accurately.}
The duration of these slices is based on the beat of the piece, as estimated by the MIDI toolbox~\citep{miditoolbox2016}. In addition, pitches are not labeled as chords. Instead, all pitches, including chord tones, non-chord tones, and ornaments, are recorded within the slice. Also, pieces are not transposed to the same key because one of the goals of this research is to explore if word2vec can learn the concept of musical key. The only musical knowledge used is octave equivalence: the pitches in each slice are converted into pitch classes. For example, both the pitch C\textsubscript{4} and C\textsubscript{5} are converted to the pitch class C. 

\begin{figure}[h]
\centering
\includegraphics[width=0.8\textwidth]{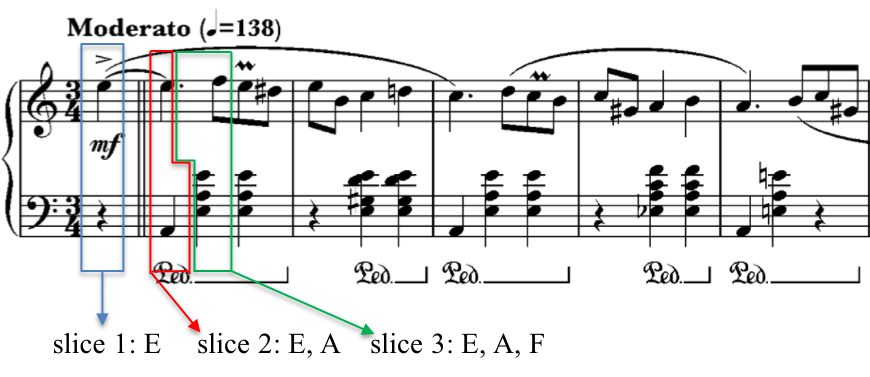}
\caption{Representing polyphonic music as a sequence of slices.}
\label{fig:slice}
\end{figure}

Figure~\ref{fig:slice} shows an example of how slices are encoded to represent a polyphonic music piece. The figure includes the first six bars of Chopin's Mazurka Op. 67 No. 4 and the first three slices for the encoding. Since a slice is a beat long, the first slice contains E, the pitch class for pitch E\textsubscript{5} in the quarter note. The second slice contains E and A because of pitches E\textsubscript{5} and A\textsubscript{3} in the second beat. Note that we include E in the second slice even though pitch E\textsubscript{5} is a tie from the first beat (not an onset) but it is still sounded in the second beat. Similarly, since the third beat contains pitches E\textsubscript{3}, A\textsubscript{3}, E\textsubscript{4}, E\textsubscript{5} (from the dotted tie), and F\textsubscript{5}, the third slice includes pitch classes E, A, and F.

The example in Figure~\ref{fig:slice} can also be used to explain the choice of beat as the slice duration. If the slice is longer than a beat, we may lose nuances in pitch and chord changes. In contrast, if the slice is shorter than a beat, we may have too many repetitive slices (where the content between slices is the same). Finding the optimal setting for the duration of the slice is out of the scope of this paper, however more research is warranted on this topic.

%\rewrite{The duration of these slices is calculated for each piece based on the distribution of time between note onsets. The smallest amount of time between consecutive onsets that occurs in more than 5\% of all cases is selected as the slice-size. The slices capture all pitches that sound in a slice: those that have their onset in the slice, and those that are played and held over the slice. The slicing process does not depend on musical concepts such as beat or time signature; instead, it is completely data-driven. Our vocabulary of words, will thus consist of a collection of musical slices. In addition, we do not label pitches as chords. All sounding pitches, including chord tones, non-chord tones, and ornaments, are all recorded in the slice. We do not reduce pitches into pitch classes either, i.e., pitches C\textsubscript{4} and C\textsubscript{5} are considered different pitches. The only musical knowledge we use is the global key, as we transpose all pieces to either C major or A minor before segmentation. This enables the functional role of pitches in tonality to stay the same across compositions, which in turn causes there to be more repeated slices over the dataset and allows the model to be better trained on less data.} 
%\dorien{I think we may need to explain the slicing a bit, given that it's a contribution of how we model the polyphonic music} \ch{will do}
In the next section, we describe a number of experiments that were conducted in order to examine whether the word2vec model can capture meaningful musical relationships.

\section{Experimental validation}

A number of experiments were set up in order to determine how well the geometry of a vector space model, built on a mixed-genre dataset of music, reflects the tonal musical characteristics of this dataset. 

%\rewrite{In order to evaluate how well the proposed model captures musical context, a few experiments were performed on a dataset consisting of Beethoven's piano sonatas. The resulting dataset consists of 70,305 words, with a total of 14,315 unique occurrences. As discussed above, word2vec models are very efficient to train. Within minutes, the model was trained on a GPU cluster... xx. }

\subsection{Dataset and model parameters}
\label{sec:dataset}
%\ch{The dataset we use is the 130,000 MIDI dataset, the so-called "the largest MIDI dataset on the Internet". See http://stoneyroads.com/2015/06/behold-the-worlds-biggest-midi-collection-on-the-internet/. I only used the songs with genre labels (to avoid some "bad" files).} 
%Download link: \url{https://mega.nz/#!Elg1TA7T!MXEZPzq9s9YObiUcMCoNQJmCbawZqzAkHzY4Ym6Gs_Q}

%\dorien{We could merge section 4.1 and 4.2? as they are rather short?}
Machine learning research that involves training models on musical corpora~\cite{boulanger2012modeling} often makes use of very specialized mono-genre datasets such as Musedata from CCARH\footnote{\url{http://musedata.org}}, JSB chorales~\cite{allan2005harmonising}, classical piano archives\footnote{\url{http://piano-midi.de}}~\cite{poliner2006discriminative}, and
Nottingham folk tune collection\footnote{\url{http://ifdo.ca/seymour/nottingham/nottingham.html}}. In this work, we aim to capture a large semantic vector space across genres. We therefore use a MIDI dataset that contains a mix of popular and classical pieces\footnote{\url{https://www.reddit.com/r/datasets/comments/3akhxy/the_largest_midi_collection_on_the_internet}}. This midi collection contains a total of around 130,000 pieces, from a total of eight different genres (classical, metal, folk, etc) and is referred to as ``the largest MIDI dataset on the Internet''\footnote{\url{http://stoneyroads.com/2015/06/behold-the-worlds-biggest-midi-collection-on-the-internet/}}. From this dataset, we used only the pieces with a genre label, in order to avoid lesser quality files. This resulted in a final dataset consisting of 23,178 pieces.%\dorien{XX} pieces. 

%\dorien{Should we some statistics about the dataset? }

%Quality of word embedding increases with higher dimensionality. But after reaching some point, marginal gain will diminish.[1] Typically, the dimensionality of the vectors is set to be between 100 and 1,000.

%\subsection{Model parameters}

\begin{figure}\centering
\includegraphics[width=.74\textwidth]{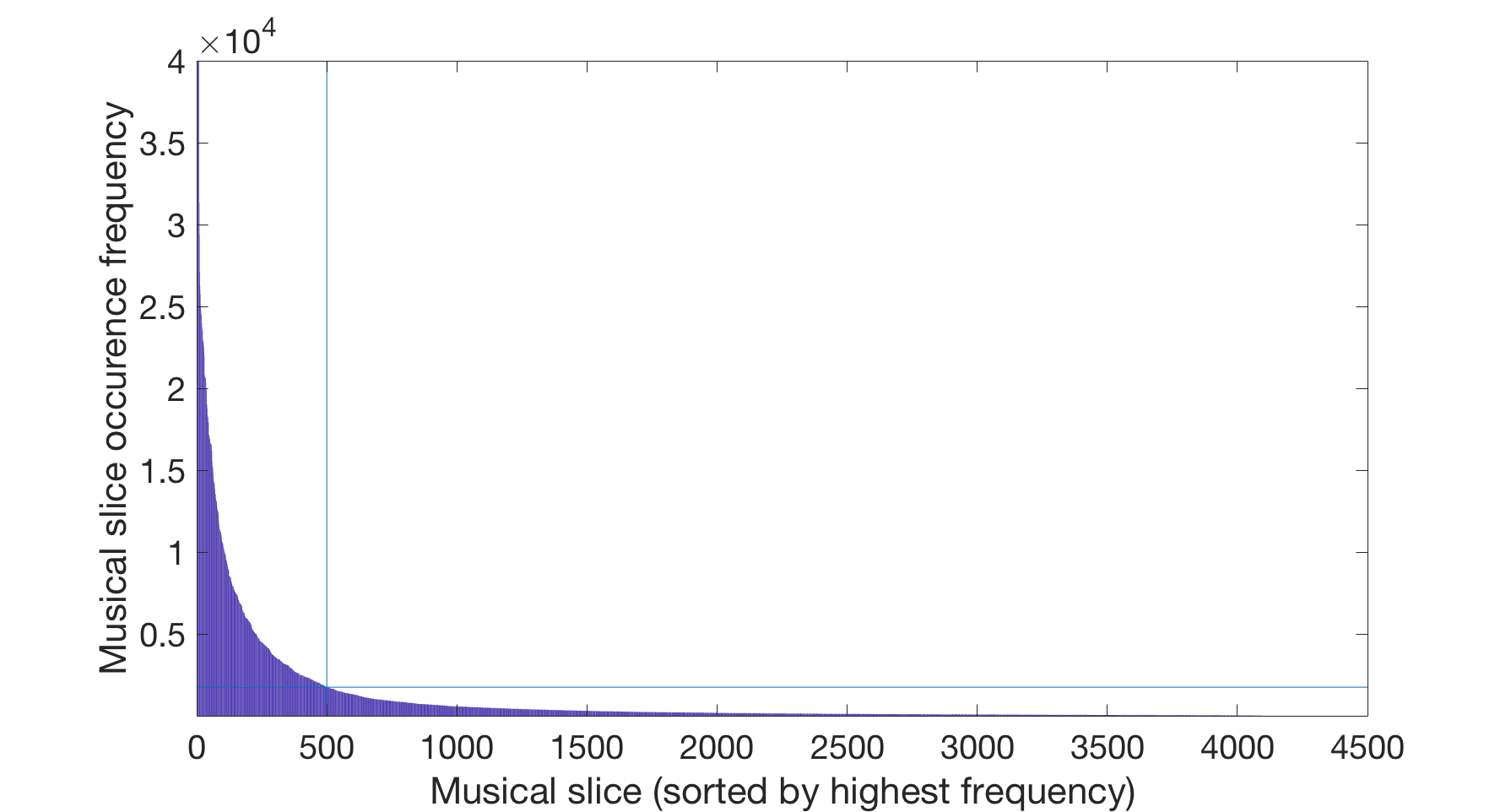}
\caption{Frequencies of each musical slice in the corpus. The blue grid lines mark the vocabulary size cutoff of 500 slices.} %\ch{Should we change the label from "word" to "slice" in Figure 3?}}\dorien{done}
\label{fig:freq}
\end{figure}

\begin{figure}[h!]
\centering
% \begin{subfigure}[h]{.44\textwidth}
% \includegraphics[width=\textwidth, height=6.4cm]{plot_training.png}
% \caption{Results for varying the number of dimensions of the vector space. }
% \label{fig:dim}
% \end{subfigure}
% \begin{subfigure}[h]{.44\textwidth}
% \includegraphics[width=\textwidth, height=6.4cm]{skipwindows.png}
% \caption{Results for varying the size of the skip window. }
% \label{fig:window}
% \end{subfigure}
% \begin{subfigure}[h]{.44\textwidth}
\includegraphics[width=.74\textwidth]{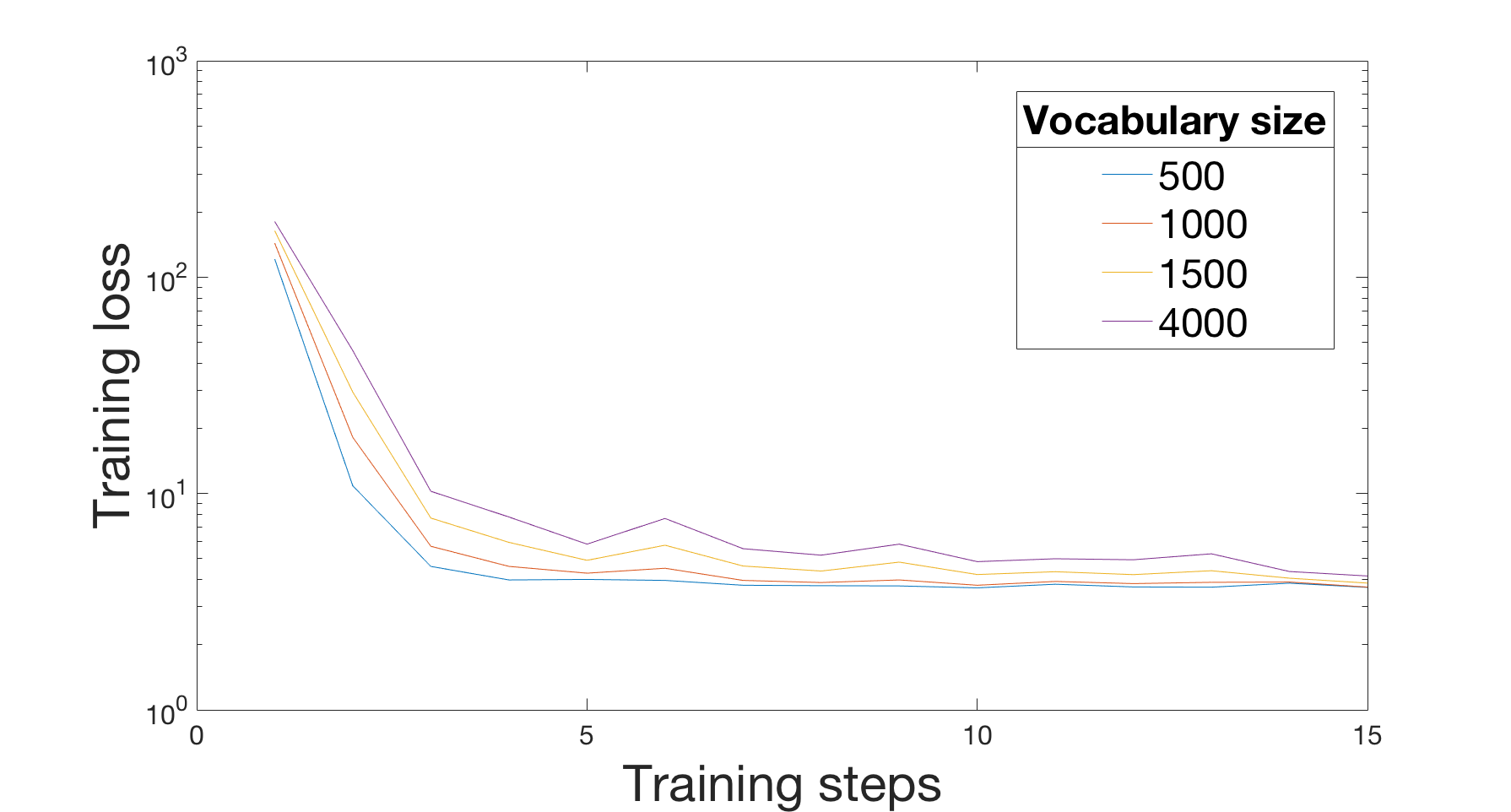}
\caption{Evolution of the average loss during training with varying vocabulary size. A step represents 2,000 training batches.}
%\caption{Results for varying the vocabulary size. }
\label{fig:vocab}
% \end{subfigure}
 %\kat{Fig needs to be updated, right?} do: updated
%\kat{I think we're deleting this figure, right? Because training loss wasn't indicative of model performance for our recent models.} }  \dorien{It is nice to have some sort of parameter tuning visualized though. We'd need to use new data though. I believe we have the data in the google doc? Or did we end up choosing a different one? More graphs may be better?}
\end{figure}

Inspired by the existing literature on NLP models~\cite{mcgregor2015distributional, pennington2014glove}, in which only the most frequently occurring word types in the text are included in the model's vocabulary, we trained our model using only the 500 most occurring musical words (slices) out of a total of 4,076 unique slices. Infrequently occurring words were replaced with a dummy word (`UNK'). This cutoff ensured that the most frequently occurring words were included, as depicted in Figure~\ref{fig:freq}. 
By reducing the number of words in our vocabulary and thus removing rare words, we were able to augment the accuracy of the model, as the included words occur multiple times in the training dataset. Figure~\ref{fig:vocab} shows that lower training losses can be achieved when reducing the vocabulary size.

%\rewrite{We trained the model a number of times, with a different number of dimensions of the vector space (see Figure~\ref{fig:dim}). The more dimensions there are, the more accurate the model becomes, however, the time to train the model also becomes longer. In the rest of the experiments, we decided to use 128 dimensions.  In a second experiment, we varied the size of the skip window, i.e., how many words to consider to the left and right of the current word in the skip-gram. The results are displayed in Figure~\ref{fig:window}, and show that a skip window of 1 is most ideal for our dataset. }

%\dorien{talk about beat, it is fixed to 1 getting the slices over all pieces} \ch{the choice of beat is described in section 3. Musical slices as words.}

A number of parameters were set using trial-and-error, including learning rate~(0.1), skip window size (4), number of training steps (1,000,000), and number of dimensions of the model (256). More details on the selection of parameters can be found in~\cite{herremans2017word2vec}.

% \begin{table}
% \begin{tabular}{cc}
%  Number of steps& \\
%  Vocabulary size& 500 \\
%  Embedding size&256\\
%  Skip window & 4\\
%  Learning rate & 0.1\\
% \end{tabular}
% \caption{Optimal parameters used in the experiment.}
% \end{table}

%experiment data on https://docs.google.com/spreadsheets/d/1VflDbKWCKBzTfCu4Y3amxgQVT23cNtNPJzOA9J2NqU4/edit?usp=sharing

%\subsection{Visualizing the semantic vector space}
\subsection{Context to concept: Chords}
\label{sec:chords}
The first musical concept that we examine is related to chords. Chords are a set of pitches that sound together and form the building blocks of harmonization. We investigate if word2vec is capable of learning the concept of harmonic relationships by examining the distance between slices that represent different chords. The goal is to see if the distance between chords in word2vec space reflects the functional roles of chords in music theory. It should be noted that, while we are studying triads in this particular experimental context, the encoding can easily handle more complex slices (from a single note to many simultaneous notes), as shown in Section~3.

In tonal music, chords are often labeled with Roman numerals in order to indicate their functional role and scale degree in a given key. For example, the C major triad (tonic) is labeled as I in the key of C major, as it is the tonic triad of the key. Triads such as G major (V, dominant), F major (IV, subdominant), and A minor (vi, relative minor) are considered common chords and closely related to the tonic triad in the key of C major. Others such as E\textsuperscript{b} major (III\textsuperscript{b}), D\textsuperscript{b} major, (II\textsuperscript{b}), and G minor (v) are less associated with I in the key of C major.  

In this study, we examined the distance between the tonic triad (I) and other triads including V, IV, vi, III\textsuperscript{b}, II\textsuperscript{b}, and v in three different keys. To calculate the distance between chords, we first mapped the chord to its pitch classes to find the corresponding musical slice. The geometrical position of this slice was then retrieved in the learned word2vec vector space. We then calculated the cosine distance between the word2vec vectors of the pair of chords. 

\begin{figure}[h!] \centering
\begin{subfigure}[h]{0.8\textwidth}
\includegraphics[width=\textwidth]{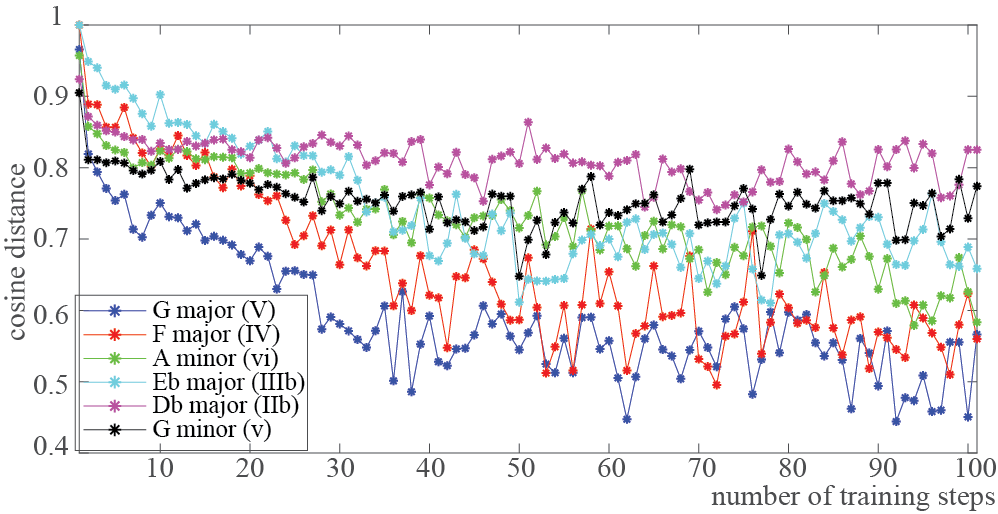}
\caption{Tonic chord = C major triad}
\end{subfigure}
\begin{subfigure}[h]{0.8\textwidth}
\includegraphics[width=\textwidth]{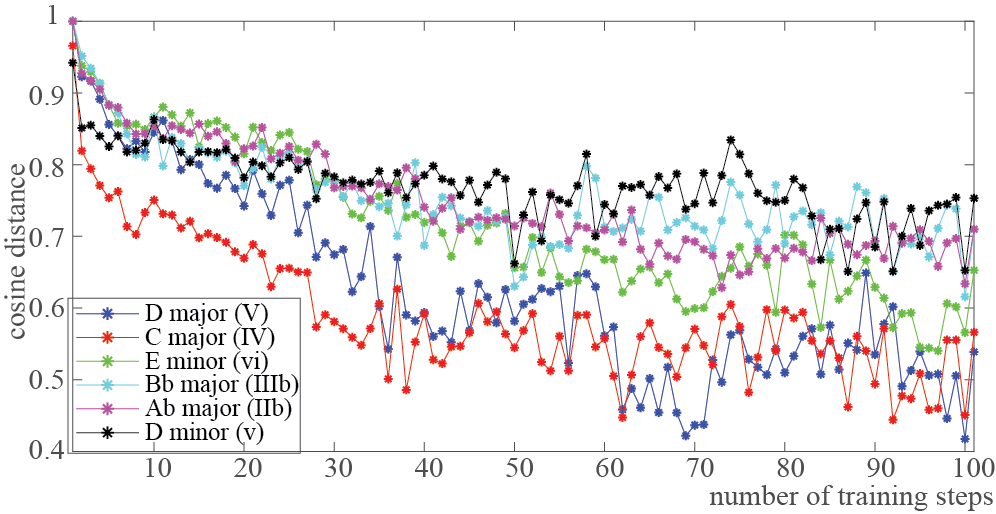}
\caption{Tonic chord = G major triad}
\end{subfigure}
\begin{subfigure}[h]{0.8\textwidth}
\includegraphics[width=\textwidth]{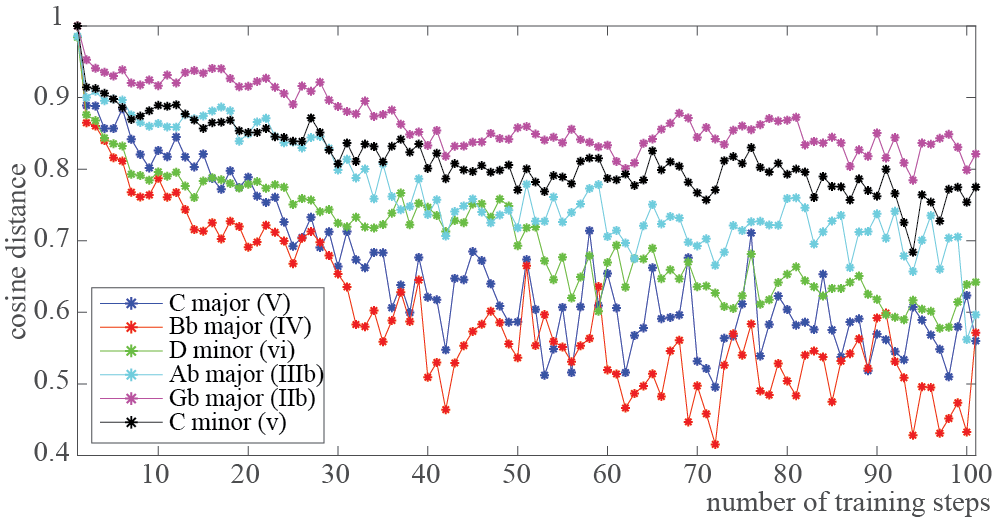}
\caption{Tonic chord = F major triad}
\end{subfigure}
\caption{Cosine distance between slices that represent a tonic chord (C, G, or F major) and its related functional chords. A training step represents 10,000 training batch examples.}
\label{fig:chord_distance}
\end{figure}

Figure~\ref{fig:chord_distance} shows the distance between (a) C major, (b) G major, and (c) F major triads. Generally, one can observe that the distances from a I triad to V, IV, and vi are smaller than those to III\textsuperscript{b}, II\textsuperscript{b}, and v. This finding is promising because it confirms the chordal relationships in tonal music theory~\cite{krumhansl1990cognitive, lerdahl1977toward}, and thus shows that word2vec embeddings are able to capture meaningful relationships between chords.

%Cosine distance between chords: Figure~\ref{fig:chord_distance}. Findings: (1). For C/G/F majors, their distance to their corresponding dominant (V) and subdominant (IV) chords is the smallest. (2) The distance to chords such as IIIb and IIb is the largest. (3) The relative minor reports the third closest distance for G major and F major, but not for C major.

%\ch{Details: 1. Train the word2vec model using all pieces in the dataset. Each piece is represented as a sequence of beats with pitch classes. 2. Once the model is trained, find the vectors for chords and calculate the cosine distance between them. 3. Focus on the distance between three chords (C major, G major, and F major) and others because C/G/F major chords are the most commonly used chords(?) We can of course test others if we want. }

%\ch{A potential question that we may get about the figures is that, instead of convergence, the result fluctuates a lot when the training iteration increases. We may want to test with different settings of parameters to see if there is one results in convergence.}

%\begin{figure}[p]
%\centering
%\includegraphics[width=0.8\textwidth]{chord_distance.png}
%\caption{Cosine distance between the to
%nic chord (a) C major, (b) G major and (c) F major, and its functional chords, in the trained word2vec space with vocabulary size = 500 and embedding size = 256. x-axis: one point represent 1000 training steps.}
%\label{fig:chord_distance}
%\end{figure}

\subsection{Context to concept: Keys}

In this section we examine whether the concept of a musical key is learned in the music word2vec space. In music theory, the relationships between keys in tonal music are represented in the circle-of-fifths~\cite{lewin1982formal}. 
In this circle, every key is a fifth apart from its adjacent neighbor, and therefore shares all pitch classes but one with the adjacent key. For example, the key of G major is adjacent to (a fifth above) C major, and contains all of the pitch classes of C major's diatonic key signature, except for including an F\textsuperscript{\#} instead of an F. When following the circle-of-fifths clockwise from a given key, the farther one is from the original key, the more different pitch classes are present. 
%\kat{Reworked this section for clarity - hope that's okay.} \ch{Looks good to me!} 
The concept of the circle-of-fifths is important in music, and has appeared in the representations or output of various machine learning models such as Restricted Boltzmann Machines~\cite{chacon2014developing}, convolutional deep models~\cite{korzeniowski2016fully}, and to some extent, RNNs~\cite{choi2016text}.
%\dorien{anybody know other refs?}\ch{will look for more}\kat{Just saw this - Yes! - Added more references, and updated text a little.} \dorien{awesome!}
In this study, we examine whether the distance between keys in the learned word2vec space of musical slices reflects the geometrical key relationships of the circle-of-fifths.
%\ch{do we need a graph of circle-of-fifth for general readers?} \dorien{It's quite basic, but if the journal is not music related, then perhaps this would be useful. I can make an image if all of us agree.}\ch{If we submit this to our special issue, we probably don't need the circle-of-fifth.} \kat{I think a written description is enough, so agreed that no graph necessary!}

We chose to experiment with Bach's Well-Tempered Clavier (WTC)'s 24 preludes, as it features a piece in each of the 12 major and 12 minor keys. The Well-Tempered Clavier is a popular music collection for computational study of tonality in music~\citep{agres2015harmonics, cancino2017bach, chew2000towards, krumhansl1990key, noland2009influences}. Each piece in the WTC was transposed to each of the other 11 major or minor keys, depending on whether the original piece is major or minor. Our augmented dataset therefore included 12 versions of the same piece, one for each key. For each piece in the resulting augmented dataset, the musical slices were mapped to the learned %\dorien{pretrained on the full corpus right? or the WTC corpus} \ch{The word2vec space is trained/built on the entire corpus. This is the only trained model and the only training done in the study.}xxx
music word2vec space (trained on the full corpus as discussed in Section~\ref{sec:dataset}). The k-means algorithm was used to identify the centroid of the slices for each piece. %\dorien{for each piece? It's not exactly clear how you find one centroid per piece with unlabeled clustering} \ch{Since each slice is mapped to a point in word2vec space, k-means just picks one point as the centroid to start and iteratively selecting different point for the centroid by minimizing the sum of the distance from all other points to the centroid} \dorien{ok, so we can put 'for each piece' right? Or could it be across pieces?}. 
We used this centroid as the %summary 
reference point for each piece in the dataset, and calculated the cosine distance between centroids as a metric of the distance between keys in the word2vec space. By comparing the centroid of a piece to the centroid of its own transposed version, one can be sure that the distance between the centroids is only affected by one factor: musical key. %\kat{I like this! Nice.}

\begin{figure}[h]
\centering
\begin{subfigure}[t]{0.49\textwidth}
\includegraphics[width=\textwidth]{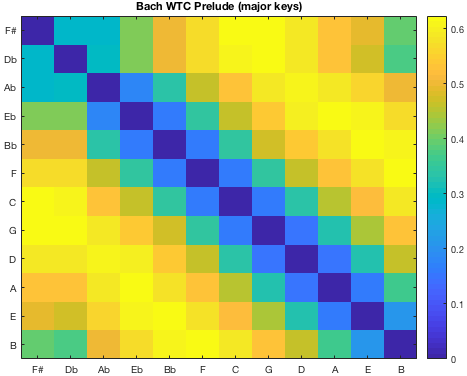}
\caption{major keys}
\end{subfigure}
\begin{subfigure}[t]{0.49\textwidth}
\includegraphics[width=\textwidth]{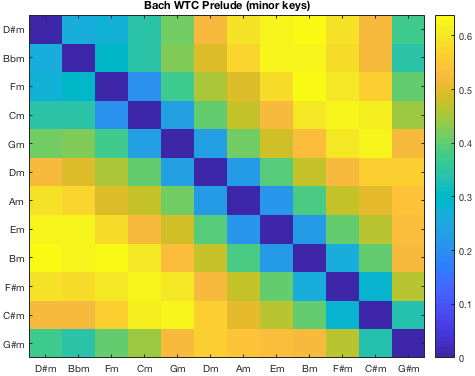}
\caption{minor keys}
\end{subfigure}
\caption{Similarity matrix for average cosine distance between pairs of each of the 24 preludes in Bach's Well-Tempered Clavier and their 11 transposed versions in word2vec embedding space.}
\label{fig:Bach_keys}
\end{figure}

Figure~\ref{fig:Bach_keys} shows the average cosine distance between %\dorien{the centroids of?}\ch{How about this?}
pairs of centroids of pieces in different keys
in the learned word2vec space. Both axes list the keys in the same order as they occur in the circle-of-fifths. Based on music theory, we expect this similarity matrix to be blue/green near the diagonal (low distance) where we find identical chords and chords %\ch{should this be keys instead of chords?} 
a fifth or two fifths apart) and red/orange (high distance) towards to the top-right and bottom-left corners, where we find pairs such as E-Db. The color should then become blue/green (low distance) again near the top-right and bottom-left corner because of the circular arrangement of keys in circle-of-fifths (e.g., F\textsuperscript{\#} is the fifth of B). It can be observed that the color patterns confirm the expectations from music theory, both in (a) major and (b) minor keys, in Figure~\ref{fig:Bach_keys}. 

%However, it does not completely reflects the circle-of-fifth. For example, the distance between key D\textsuperscript{b} is consistently large to any other keys. This is because D\textsuperscript{b} is a less common key used in (popular) music \ch{I want to cite the blog Kat found about commonly keys/chords. Is it "academic" enough? http://www.hooktheory.com/blog/i-analyzed-the-chords-of-1300-popular-songs-for-patterns-this-is-what-i-found/}  \dorien{Perhaps we can refer indirectly: as shown from chord statistics based on the popular music dataset available at xx?}

%\ch{Perhaps we can transpose all Bach's WTC 24 pieces to all 12 keys, and calculate the overall average of the distances between pairs of keys so that we might be able to do some statistical analysis.}

\subsection{`Analogy' in music}
One of the most striking results from word2vec in natural language processing is the emergence of analogies between words. \citet{mikolov2013linguistic} studied linguistic regularity for examples of syntactic analogy such as `man is to woman as king is to queen'. They proposed the analogy question to the learned word2vec space as $xa:xb = xc:xd$, where $xa$, $xb$, $xc$ are given words. Their goal was to search for $xd$ in the word2vec space where the vector $xc$-to-$xd$ is parallel to the vector $xa$-to-$xb$. %\dorien{just checking: a word is a vector in word2vec. So we are talking about a vector between vectors? Would it be accurate to write $\vec{xa}$ (sorry overhead arrow not showing for some reason), or just $w_i$ for each word?} \ch{xa is a point in the word2vec multi-dimensional space. Because of the multi-dimension, it is easier to call it a vector with n elements for a point (but it is not the vector we know as in linear algebra since it is actually a point). However, from xa to xb (ex. from the word `man' to `woman') is a vector. I use xa/xb... is because that's what the authors in \citep{mikolov2013linguistic} used} DH: got it!
The authors found $xd$ by searching for the word for which $xc$-to-$xd$ has the highest cosine similarity %\dorien{Aren't we using cosine distance as a metric?} \ch{Yes, we are. But I was just describing what the authors in \citep{mikolov2013linguistic} did for analogy in language} DH: sorry my bad!
to the vector $xa$-to-$xb$. 

Although music lacks the explicit semantic content of language, researchers have argued that the construct of analogy exists in music as, for example, the ``functional repetition'' of musical content 
that is otherwise dissimilar~\cite{kielian1990interpreting}. 
%, drawing upon organizational elements to impact the perception of the listener
%\kat{Hope this is okay.}\ch{It's great!} \kat{phew! ;)}
%\kat{[Note to self: Consider the meaning of analogy in music as opposed to language... Whereas two vectors with the same orientation (parallel vectors) have a cosine similarity of 1, we would not expect CosSim of 1 for meaningful chord/key relationships. Tie this in with circle of fifths..]} \ch{no more CosSim, just degree now:)}
Therefore, inspired by computational linguistics~\citep{mikolov2013linguistic}, we explore the meaning of analogy in the music word2vec space. 
Perhaps the most fundamental definition of analogy in tonal music would be the functional role of chords in a given key. For example, the relationship between the tonic chord and the dominant (I-V) should remain the same in all keys, i.e., C major triad is to G major triad in the key of C major as G major triad is to D major triad in the key of G major. Based on this definition, we investigate the analogy or relationship between a vector representing the transition from one chord to another (called a 'chord-pair vector'), and its transposed equivalent in every key. 
%and this chord-pair vector's transposed equivalent chord-pair vectors. 

For the present experiment, we focused on three chord-pair vectors: C major-to-G major (I-V chord-pair vector in major keys), A minor-to-E minor (i-v in minor keys), and C major-to-A minor (I-vi, the tonic and its relative minor triad, in major keys). 
In order to accommodate working with music instead of language, we had to slightly adapt the question of how analogy is represented geometrically: instead of searching for $xd$ given $xa$, $xb$, and $xc$, we calculated the angle between the two chord-pair vectors $xa$-to-$xb$ and $xc$-to-$xd$ given all four variables (chords). This change was made because a vector in music word2vec space may not necessary be a chord, e.g., it could  contain only one or two pitch classes. %\dorien{do you mean because a slice can contain only two note for instance? I'm not sure we need to highlight that we 'change' the analogy question } \kat{two notes is less common but still a chord.. but slices can contain only 1 note, right? also, some slices contain many notes that don't translate to a normal chord like GM7, etc}. \ch{A slice can contain anything, one note, two notes, even five notes. The problem with the language approach is that I can find something for xd like a note B, or a slice of five notes (D,E,F,A,B) for xa=C major triad, xb=G major triad, and xc=G major triad} \dorien{that makes sense. I'm just wondering what the change is then 'given all four variables are present?', currently it sounds liek the change is that we are looking at the chord-pair vectors instead of absolute positions?}
If the retrieved $xd$ is not a chord that is commonly recognized by music theory, we cannot verify whether it is a meaningful analogy. 

\begin{figure}[h!] \centering
\begin{subfigure}[t]{0.55\textwidth}
\includegraphics[width=\textwidth]{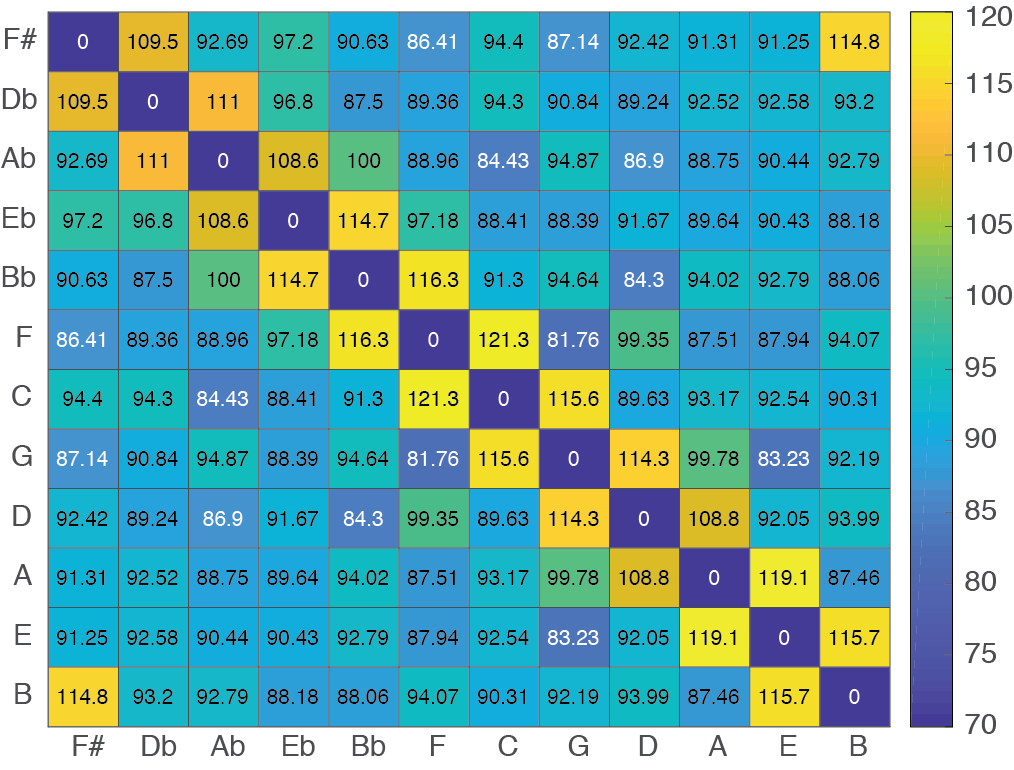}
\caption{}
%\caption{Angles in degree between pairs of I-V vectors in different keys.}
\label{fig:vector_C_G_a}
\end{subfigure}
\begin{subfigure}[t]{0.35\textwidth}
\includegraphics[width=\textwidth]{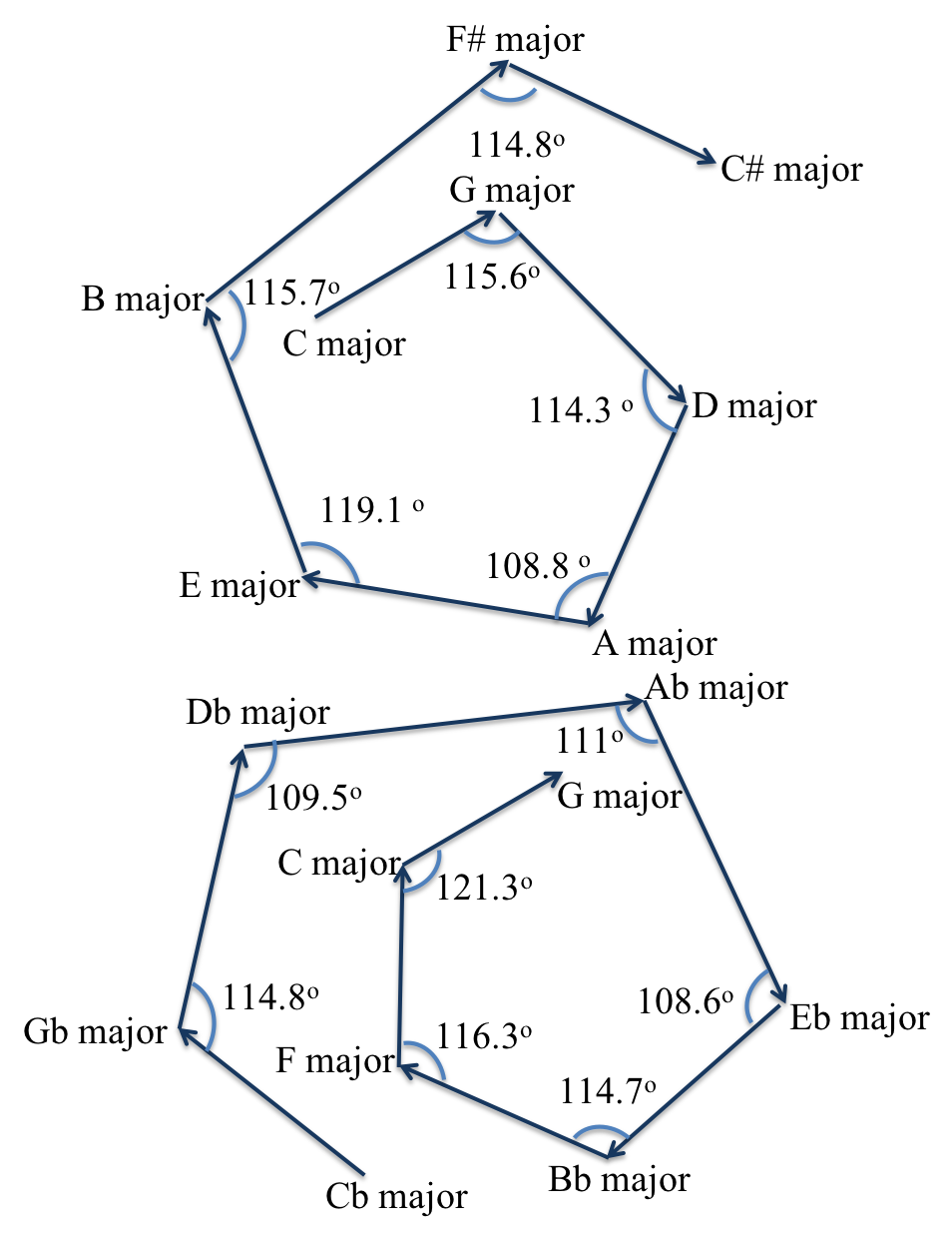}
\caption{}
%\caption{Illustration of chord vectors and angles}
\label{fig:vector_C_G_b}
\end{subfigure}
\caption{(a) Similarity matrix for the angle (in degrees) between I-V vectors in pairs of keys (keys are indicated by the x-and y-axes) and (b) an illustration of the chord-pair vectors and angles between them. Note that the angles between chord-pair vectors are computed between adjacent chords; that is, the reader should not draw conclusions about the relationship between non-adjacent chords from this diagram. In addition, it should be noted that the lengths of the arrows were chosen to maximize the clarity of the figure, and as such do not convey semantic meaning in of themselves.
%\ch{In addition, the length of the arrow is determined based on the clarity of the illustration and does not imply any semantic meanings.} \dorien{Perhaps this is more clear?: In addition it should be noted that the the length of the arrows was chosen to maximize the clarity of the figure, and as such it does not have any semantic meaning. }
%\kat{I still felt that a disclaimer was needed, so I added one. Hope that's okay.}\ch{I am okay with the disclaimer.} 
%\kat{These are very cool, but I have a couple of questions. First, why are the cells around the diagonal \textit{negative} CosSim? Fig b is quite clever :) I'm just wondering, to generate this graph, you computed Bb to F, and F to C, etc, right? But do the other relationships hold, e.g., from Bb to C, etc? I just want to make sure this graph isn't misleading in terms of all of the chord relationships.}
}
\label{fig:vector_C_G}
\end{figure}

%\dorien{instead of chord vector, could we call it chord transition vector? I think that may make it more clear.}\kat{chord pair? transition implies something functional in a music theoretic sense.}\ch{Just to clarify, a chord vector is a vector (as defined in linear algebra) from one chord to another in word2vec space. C major triad is a point, G major triad is another point. From C major triad to G major triad becomes a vector. I guess in this case we can call it a chord pair vector. What do you think?} \dorien{agreed that chord pair vector might be more clear as we are referring to single slices as vectors sometimes. }
Figure~\ref{fig:vector_C_G_a} lists the angle in degrees between the I-V chord vectors in pairs of keys (where the x-and y-axis indicate every key).

%\begin{figure}[h!] \centering
%\begin{subfigure}[h]{.44\textwidth}
%\includegraphics[width=\textwidth]{tsne_E_Eb.png}
%\caption{E (green) and Eb (blue).}
%\label{fig:vis1}
%\end{subfigure}
%\begin{subfigure}[h]{.44\textwidth}
%\includegraphics[width=\textwidth]{tsne_Eb_Db_B.png}
%\caption{Eb (black), Db (green) and B (gray).}
%\label{fig:vis2}
%\end{subfigure}

%\dorien{which keys?} \kat{C and G, right?}. \ch{The labels are keys. If you look at the cell in row G and column C, this means the degree between the chord pair vector I-V in G major key and the chord pair vector I-V in C major key. In this case, it is the angle between the chord pair vector G maj triad-to-D maj triad and the chord pair vector C maj triad-to-G maj triad } 
For example, the value 121.3 in row C and column F is the degree between the chord-pair vector C major-to-G major (I-V in C major) and the chord-pair vector F major-to-C major (I-V in F major). %\dorien{This is not clear to me. Does the row go to C major and the column to Fmajor? Is it not the angle between C and F chord? }. 
It is apparent that the numbers next to the diagonal line, which represent keys a fifth apart, are significantly different from other numbers in the matrix. 
To help the reader interpret the meaning of these highlighted numbers, we illustrate the I-V chord-pair vectors with their angle between different keys in Figure~\ref{fig:vector_C_G_b}. Note that Figure~\ref{fig:vector_C_G_b} almost mimics the circle-of-fifths, although it does not perfectly fit all 12 keys in the projected 2D circle due to the multidimensional nature of the underlying vectors.
%\kat{updated this section for clarity as well.}\ch{Looks good.}
This again confirms that the relationships between keys in the circle-of-fifths is reflected in the learned vector space, as the angles between I-V chord-pair vectors all fall within a very specific range (108.8-121.3 degrees). 

%\dorien{Could we phrase it positively and say that it fits 8 out of 12 keys? Is that correct to say? } \ch{Probably not. In circle of fifth, since there are 12 keys in a circle, essentially we are looking at a 12-sided polygon (dodecagon?). The internal angle between adjacent vertice in a dodecagon is 150 degree. What we found in word2vec is around 120 degree. But again, circle-of-fifth is 2D and word2vec is multi-dimensional. At least we can show that the degree between I-V chord pair vectors in adjacent keys is significantly and consistently different from other key pairs.} DH: good point!
%Given, the original word2vec space is multi-dimensional and Figure~\ref{fig:vector_C_G_b} is only a projection to 2D space. %\dorien{we aren't doing this right? we could say this in future research.}\kat{Agreed, this would fit in nicely with what I was writing for the future work section. Should I move that last sentence starting `Therefore...' to the Conclusion?}\ch{Yes, this can be the future work.}

\begin{figure}[h!] \centering
\begin{subfigure}[t]{0.48\textwidth}
\includegraphics[width=\textwidth]{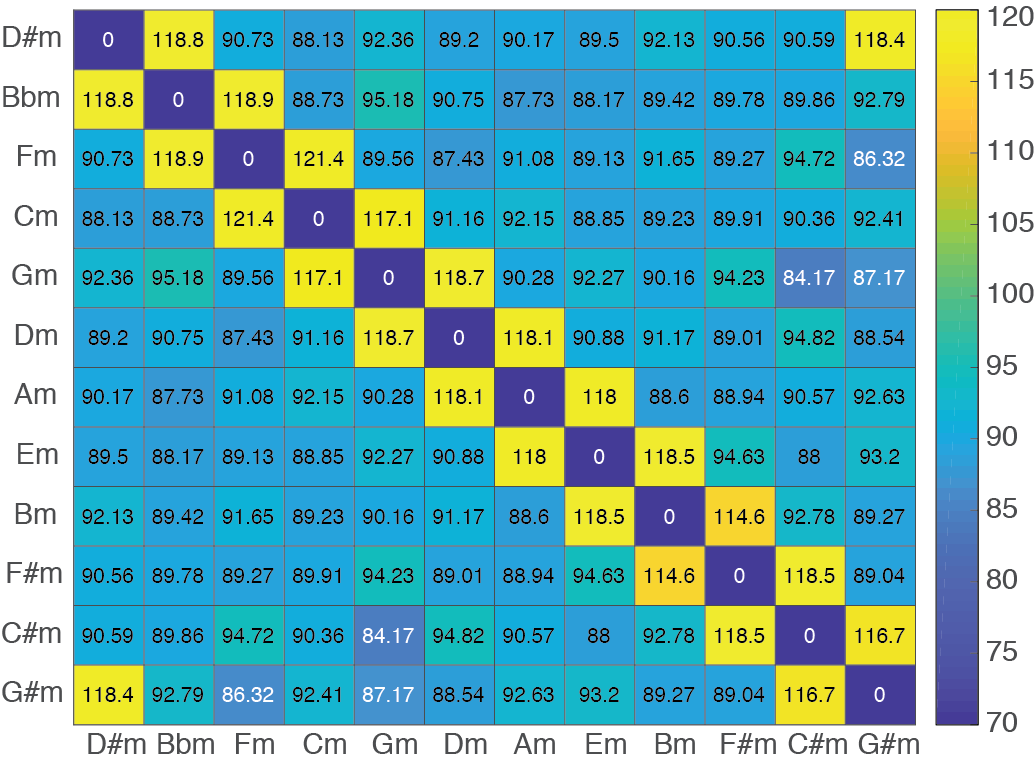}
\caption{i-v}
\label{fig:vector_Am_Em}
\end{subfigure}
\begin{subfigure}[t]{0.475\textwidth}
\includegraphics[width=\textwidth]{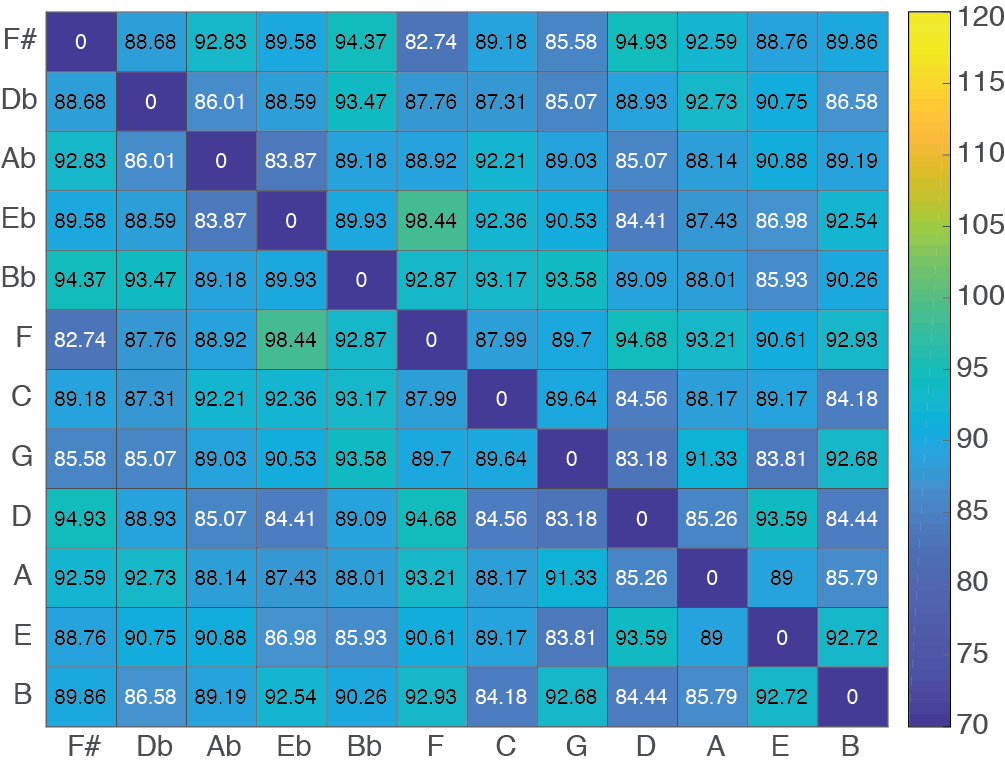}
\caption{I-vi}
\label{fig:vector_C_Am}
\end{subfigure}
\caption{Similarity matrix for the angle (in degrees) between chord-pair vectors, (a) i-v and (b) I-vi, respectively, for pairs of minor keys (a) and for pairs of major keys (b). Keys are listed along the x-and y-axes.}
\label{fig:vector_others}
\end{figure}

Figure~\ref{fig:vector_others} shows the similarity matrices for the angle between chord-pair vectors i-v in pairs of minor keys 
%(as indicated by the x-axis and y-axis in Figure~\ref{fig:vector_Am_Em}) 
and I-vi in pairs of major keys.
%(as indicated by the x-axis and y-axis in Figure~\ref{fig:vector_C_Am}). 
One may observe that Figure~\ref{fig:vector_Am_Em} also shows a different pattern along the diagonal line, implying that the circle-of-fifths relationship is also learned in the music word2vec space for minor keys. In contrast, the relationship between the tonic and its relative minor (I-vi) is not maintained, as no significant patterns are observed in Figure~\ref{fig:vector_C_Am}. It might be that the I-vi chord-pair vector does not occur as frequently as the I-V vector or that the context in which the I-vi chord-pair vector appears tends to be more diverse. %This is to be expected, given that i-vi chords are  %\kat{Revised.}\ch{looks good.}

\subsection{Music generation with word2vec slices}

\begin{algorithm}
\SetKwInOut{Input}{Input}
\SetKwInOut{Output}{Output}
\SetKwInOut{Parameter}{Parameter}
%\underline{function searchSlice} $(s,n, V)$\

\Input{$s$: input slice, $W2V_s$: slices in the word2vec vocabulary with their embeddings}
\Output{$\overline{s}$: the substitute slice for $s$}
\Parameter{$n$: top $n$ candidate slices}
\If{$s$ is not in $W2V_s$}{
$\overline{s} \leftarrow s$\;}
\Else{
  \emph{// initializing arrays for the slice, cosine distance, pitch class score, and pitch class count of the top $n$ candidates (with 1 representing the same number of pitch classes compared to original piece and 0 representing a different amount) \;}
  \For{$i$ = 1...$n$}{$slice_n[i]\leftarrow$Inf\; $distance_n[i]\leftarrow$Inf\;
  $score_n[i]\leftarrow$0\;
  $count_n[i]\leftarrow$0\;}
  \emph{// finding the top/closest n slices to s\;}
  \ForEach{slice $t$ in $W2V_s$}{
    \If{cosdis($s$,$t$)$<$max($distance_n$)}{
      $i\leftarrow argmax(distance_n)$\;
      $slice_n[i]\leftarrow t$\;
      $distance_n[i]\leftarrow$ cosdis($s$, $t$)\;
    }
  }
  \emph{// calculating weights for the 12 pitch classes in array pitchclasses\;}
  \ForEach{pc in pitchclasses}{$pc\leftarrow$0\;}
  \ForEach{slice $t$ in $slice_n$}{
    \ForEach{$note$ in $t$}{
      $pitchclasses[note]\leftarrow pitchclasses[note]$+1
    }
  }
  \ForEach{pc in pitchclasses}{
    $pc\leftarrow pc$/sum($pc$)
  }
  \emph{// selecting $\overline{s}$\;}
  \For{$i$=1...$n$}{
    \ForEach{$note$ in $slice_n[i]$} {
        $score_n$[$i$] = $score_n$[$i$]+$pitchclasses[note]$
    }
    $score_n$[$i$] = $score_n$[$i$]/(total number of pitch classes in $slice_n$[$i$]\;
    \If{number of pitch classes in $slice_n$[$i$] == number of pitch classes in $s$}{  $count_n$[$i$] = 1; 
    }
  }
  \emph{// when no slices with same number of pitch classes are in top-n list\;}
  \If{sum($count_n$)==0}{
    $i\leftarrow$ argmax($score_n$)\;
    $\overline{s}\leftarrow slice_n$[$i$]\;
  }\emph{// when there are slices with same number of pitch classes in the top-n list\;}
  \Else{
   %$i\leftarrow$ argmin($distance_n$)\;
   % $\overline{s}\leftarrow slice_n$[$i$]\;
  %\emph{//DH: I THINK THE BELOW CAN BE DELETED THEN\;}
    $m\leftarrow$0\;
    $j\leftarrow$0\;
    \For{$i$=1...$n$}{
      \If{$count_n$[$i$]==1 and $score_n$[$i$] $>m$}{ 
        $m\leftarrow score_n[i]$\;
        $j\leftarrow i$\;
      }
    }
    $\overline{s}\leftarrow$ $slice_n$[$j$]\;
  }
  
}
\caption{Finding the substitute slice for music generation}
\label{algo:algorithm}
\end{algorithm}

Because we have shown that word2vec is capable of recognizing different relationships between chords in Section~\ref{sec:chords}, we give an example of a potential application of word2vec for music: music generation by means of substitutions. Systems that generate music by recommending alternative chords or pitches can be useful for composers and amateurs to experiment with new ideas or that are not in their repertoire. Music generation systems are also becoming popular to create music for commercial applications (for example, see Jukedeck\footnote{www.jukedeck.com/}). For a complete overview of current state-of-the-art of music generation systems, the reader is referred to~\citet{herremans2017functional}. 

%\dorien{There are some page layout issues. How about putting the algorithms in appendix?} \kat{Hmm, nicer to have them in the body of the text though, no? They're not really supplemental material here.}\ch{Usually the algorithm is with the text, but I can see the layout issue. Is it because we have too many figures (than the text)? If so, I can try to make Fig7(a) and (b) side-by-side and Fig8(a) and (b) side-by-side.} \dorien{thanks! yeah, it's better now :)} 
%\kat{My only comment is that now Fig 7B is hard to see, and the numbers in Fig 8A and B are also very hard to read. UPDATE: I made Fig 7B larger, so now it's easier to read, but not sure what to do about Fig 8.} \ch{Thanks for making Fig 7B bigger! For Fig 8, the only thing I can think of is to round the numbers to integer and hopefully less content means they can be presented in a bigger font. If not, at least we have the color bar to show the range of the value.} \dorien{Yeah maybe it's ok now. We can see if the publisher objects :)? \kat{okay}}

%\dorien{Thanks CH for your clarifications of the algorithm! I made one clarification and one modification, see capital letters in the algorithm.  } \ch{Thanks a lot! The first clarification in the comment is great! I keep lines 49-57 the same because the algorithm uses the highest pitch class score, not the shortest cosine distance. I mistakenly wrote shortest distance last time. Sorry!} \dorien{got it, ok!}

Algorithm ~\ref{algo:algorithm} describes the strategy that we propose for finding a substitute slice for a given input slice using the trained word2vec model. Note that the pitches in a slice may or may not represent a chord. If the input slice is not in the word2vec vocabulary, the algorithm returns the original input slice as its own substitute (lines 1-3). Otherwise, it calculates the cosine distance between all slices in the vocabulary and the input slice (lines 6-11). It then collects a list of \textit{n} slices that are the closest to the input, where \textit{n} is an input parameter of the algorithm (lines 13-19). Because the music considered in this study is tonal, the algorithm uses the top-\textit{n} slices to  calculate a score for the 12 pitch classes, which can be used to infer the key. %\dorien{I'm curious if this key is the same as the key of the original fragment?}  %\dorien{is this musical key used for something? I don't see it reoccur?} 
%\ch{I didn't check the key to see if it is the same as the original.
%\ch{As we discussed, we used the pitch class score to select the ones that occur more often in the top-n slices. And based on what we observed, word2vec recognizes the distance between keys, so slices in the same region may belong to the same key. As in most of the key finding algorithms, pitch class distribution is used to identify the key. Therefore, I only used pitch class distribution but not identify the key explicitly.} \dorien{Got it, so basically, if C is the most occurring pitch class, we say it's C major} 
The score of a particular pitch class is its normalized frequency of occurrence in the top-\textit{n} slices (lines 21-41). 

As a design rule, we chose to preferentially substitute a slice with one containing the same number of pitch classes. Therefore, we added a preference rule in the algorithm: From the top-\textit{n} slices, the algorithm selects the slice with the highest pitch class scores %\kat{This confuses me:} [to the input slice] \kat{can we delete these three words?} 
among those with the same number of pitch classes. %\ch{Sorry for the confusion. When selecting the slice, I always select the one with the highest pitch class score. The cosine distance is only used to select the top-n slices. In this way, we will have difference slices when the value of n changes. If I use the cosine distance, the selected slice will be the same as long as there are no slices with the same number of pitch classes.} \dorien{Great, I understand now.}. 
If none of the slices in the top-\textit{n} list have the same number of pitch classes as in the input slice, the algorithm returns the slice with the highest pitch class score regardless of the number of pitch classes (lines 38-40 and 42-57). %\dorien{Don't we use the information about the pitch class frequency?}

To show the effectiveness of the algorithm, we applied the algorithm to Chopin's Mazurka Op. 67 No. 4 and generated substitute slices for the first 30 beats. For the input parameter \textit{n}, we experimented with four settings for the top-\textit{n} list: 1, 5, 10 and 20 slices. The details of the generated slices and their cosine distance to their respective input slice can be found in Figure~\ref{fig:chopin_details} in Appendix. 

\begin{figure}[h] \centering
\includegraphics[width=0.85\textwidth]{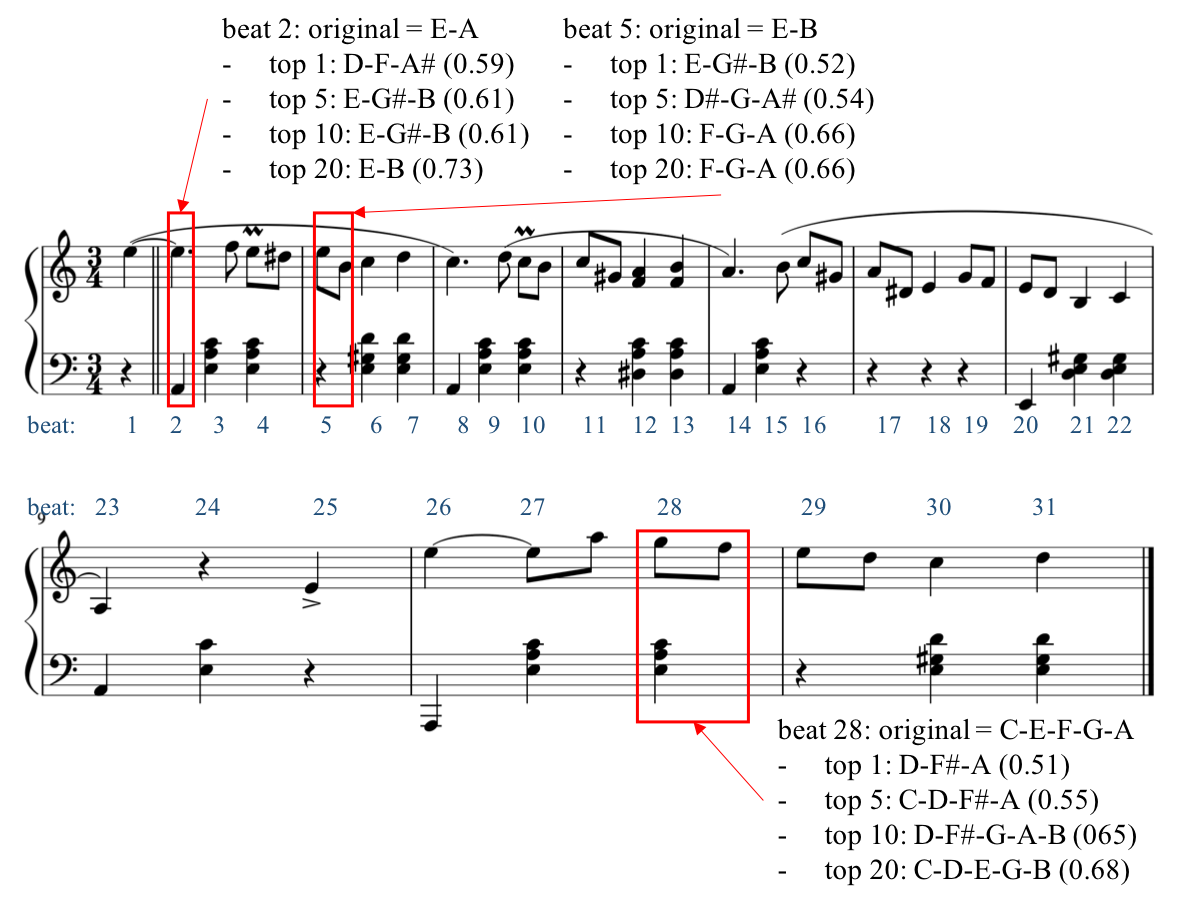}
\caption{Chopin Mazurka Op. 67 No. 4, with newly generated slices for three selected beats. Each of these slices is annotated with the cosine distance %\ch{It is cosine distance. We don't use cosine similarity anymore} 
of the best slice (according to Algorithm~\ref{algo:algorithm}) in a set of top-n lists. %\dorien{CH: I am correct right? Or is this just the 1-5-10th closest slice?} \ch{For each n in the top-n setting, the best slice is shown here for the selected beats. For example, on beat 2, the slice for top 20 contains pitch classes E and B, and its cosine distance to the original (which has pitch classes E and A) is 0.73.} \dorien{Got it, than I understood correctly, just wanted to verify}%\kat{I'm sort of wondering why/how you chose these three beats}\ch{shorter distance (roughly the local minima in Fig~\ref{fig:chopin_details})}
}
\label{fig:chopin_score}
\end{figure}

To illustrate the effect of using different values for \textit{n}, we selected the beats for which the substitute slice has a relatively short distance to the original, and then depict them on the score, as shown in Figure~\ref{fig:chopin_score}. %\kat{any justification for this choice?} \ch{how about this:} \kat{Thanks, this reads better now!}
Three beats were selected: beats 2, 5, and 28. As shown in Figure~\ref{fig:chopin_details}, the substitute slice for these three beats has a smaller distance to the original compared to the other beats. In addition, the slice that the model selects for substitution in the selected beats changes as \textit{n} increases.
For each of the three highlighted beats in~\ref{fig:chopin_score}, we annotated the pitch classes in the original slice (as shown in the score) and of the substitute slice, and calculated the cosine distance between them (in parentheses). 
The most notable trend in Figure~\ref{fig:chopin_score} is that the number of out-of-key pitch classes decreases when the value of \textit{n} increases. Because this piece is in the key of A minor, pitch classes with sharps (\#) or flats (\textit{b}) are considered to be outside of the key. When \textit{n} = 20, none of the selected slices contain any sharps or flats, which is surprising, as the cosine distance also increases when the value of \textit{n} increases. 
%\dorien{The system performs in this way due to the fact that it is able to learn information about keys.} %\ch{Dorien and Kat: please look at the figures~\ref{fig:chopin_score} and ~\ref{fig:chopin_details} to see if there is anything else that you want to add here.} \kat{I like Fig 9! And agree that Fig 10 might be best in the appendix - it's interesting, but very busy and a little hard to read.} \dorien{I've put it in appendix}\kat{Thanks! Can we force Fig 9 to appear before the Conclusion?}\ch{Thanks!}

Readers may listen to examples of the generated pieces online\footnote{http://sites.google.com/view/chinghuachuan/}. It should be noted that our aim was not to create a full-fledged music generation system, but rather, to illustrate how word2vec might be useful in a music generation context. 
Although we have only described one method for slice replacement above, there are numerous ways in which one might appropriate word2vec for slice/chord replacement in music. In future research, it would be interesting to explore how word embeddings, which, as we have demonstrated, capture meaningful musical features, may be used as input for other music generation models.

\section{Conclusion}

%\kat{[Hi both, I've revised the Conclusion a bit. I think we're about done now, with the possible exception of combining the third and fourth paragraphs?]} \dorien{I think it's good as it is} \kat{Okay, if the part below on analogy is good for everyone, then I think the conclusion is finished!} \dorien{ok for me!}

In this paper, we explore whether a popular technique from computational linguistics, word2vec, is useful for modeling music. More specifically, we implement a skip-gram model with negative sampling to construct a semantic vector space for complex polyphonic musical slices. 
We expand upon preliminary research from the authors~\citep{herremans2017word2vec}, and provide the first thorough %, preliminary 
examination of the kinds of semantic information that word2vec is capable of capturing in music. 

In our experiment, we trained a model on a large MIDI dataset consisting of multiple genres of polyphonic music. This allowed our dataset to be much larger and more ecologically valid than what is traditionally used for music-based models. 
%Although some previous preliminary research [made anonymous for submission] \cite{huang2016chordripple} has employed word2vec to extract harmonic relationships, the present work does not start by giving the model explicit musical information, such as chord labels, key information, or any other type information about musical relationships. 
%\dorien{Not sure if this sentence makes sense in a word2vec context, because words are just treated as labels. The fact that Anna used F and Fm, would not have changed her model if she labeled them A and B. }
%\dorien{MAY I SUGGEST THIS INSTEAD: 
Although previous preliminary research \cite{huang2016chordripple} explored the use of word2vec by modeling 92 labeled chords, the present work builds upon initial work by the authors~\citep{herremans2017word2vec} by using a vocabulary of 4,076 complex polyphonic slices with no labels. In contrast to many traditional music modeling approaches, such as convolutional neural networks using a `piano roll' representation, our representation does not encode any musical information (e.g., intervals) within the slices. We sought to explore whether word2vec is powerful enough to derive tonal and harmonic properties based solely on the co-occurrence statistics of musical slices. 
%\kat{(Have updated this.)} \dorien{Super.}\ch{The conclusion looks good to me!}
%} \ch{Looks good.}

% Huang: 92 unique chord symbols
First, we found that tonal distance between chords is indeed reflected in the learned vector space. For instance, as would be expected from music theory, the tonic and dominant chords of a key (I and V) have smaller cosine distance than the tonic and the second (I and II) chords. This shows that elements of tonal proximity are captured by the model, which is striking, because the model received no explicit musical information about the \textit{semantic content} of chords.
% Second, we observed that the relationships between keys are reflected in our geometrical model: 
%\kat{I'm confused... why are the results about `the angle between I-V chord-pairs' listed second here? It should be last. We discuss Figs 7 and 8 before Fig 6. I'm changing the order we talk about the results.. this will also allow for a smoother transition into the next paragraph.}
Second, %\kat{New order:}
we observed that the relationships between keys are reflected in our geometrical model: by transforming Bach's Well Tempered Clavier to all keys, we were able to verify that the cosine distance between the centroids of tonally-similar keys is much smaller than for tonally-distant keys.
%\ch{Kat, you are talking about Figure~\ref{fig:Bach_keys} or Fig ~\ref{fig:vector_C_G}? \kat{I was referring to Fig 7, assuming that the other keys weren't plotted because the nice circle-of-fifths relationships fell away..} The color in Figure~\ref{fig:Bach_keys} shows the cosine distance, not angle (Fig ~\ref{fig:vector_C_G} is angle). Are you referring to Fig ~\ref{fig:vector_C_G} regarding the 8 out of 12 keys in the analogy section? Just to clarify, we can include all 12 keys in Fig~\ref{fig:vector_C_G_b} as the rest four keys also have similar degrees around 110 to 120. \kat{Ah! I didn't realize that the circle of fifths (angles between keys) continued! It would be great to express that all of the angles between keys a fifth apart roughly mirror the CoF.}
%\ch{It is just a bit tricky to put all 12 in 2D. If you want (and to avoid confusion), I can add the rest four keys to Fig~\ref{fig:vector_C_G_b}} \kat{How about showing the COF separately for sharp keys and flat keys? We can't make a complete circle with all 12 keys anyway. Maybe we could make one figure for C G D A E B F\# (C\#), and another for C F Bb Eb Ab Db Gb (Cb). (And yes, I realize F\# and Gb are the same centroid, but that's okay I think.)  Would that work?} \ch{Done. See if the new figure works.} \dorien{I think it looks great and captures the multidimensionality quite well. Thanks for this!}
Third, we observed that the angle between I-V chord vectors between pairs of keys (e.g., G major to D major, Eb major to Bb major, etc) in vector space is generally consistent with the circle-of-fifths. This suggests that the model was able to extract fundamental harmonic relationships from the slices of music. 
%\ch{Is this (previous two sentences) about Figs. 7 and 8? Maybe we can move the second point to the next paragraph where we talk about musical analogy, so we can have this paragraph focuses on chords and keys (Figs 5 and 6).} \kat{Okay, I think the point of this paragraph is just to give a quick overview of all the findings. I'd keep the three points here but change the text below.}
%Transcribing the Bach dataset to all keys allowed for key relationships to be discovered; it is possible that a circle-of-fifths relationship for the other four keys might be discovered by augmenting this dataset (especially with additional music in the key of E major, B major, F\# major, and C\# major). 
In sum, not only are tonal and harmonic relationships learned by our model over the course of training, but the associations between keys are learned as well.

%\kat{The paragraph above (starting "First, we found...") sort of summarizes the main findings, so the paragraph below above seems a little redundant (except for mentioning `analogy'). I'm wondering if we should fold it into the paragraph above. Do you think so, or is it okay as it is?} \ch{This paragraph is supposed to talk about musical analogy, right? Maybe add a sentence to explain the definition of musical analogy. See if the following is good.} \kat{Good idea. I've updated the definition - is it okay?} \ch{Yes, it looks good!}

By extracting information about chords and chord relationships, the model's learn-ed representations highlight the building blocks of musical \textit{analogy}. In the present approach, musical \textit{analogy} is defined as the semantic invariance of chord transitions, such as I-V, across different keys. That is, similar translations (e.g., the angle between chord-pair vectors) can be used to move between functional chords in different keys because the angle between chord pairs (such as I-V) is preserved regardless of the key. 
%That is, when the relationship between functional chords is preserved regardless of the key. \dorien{I thought analogy was defined as similar translations necessary to move between functional chords in different keys (e.g. the angle), but perhaps the previous is more general, so then you can ignore my comment.} \kat{What if we combine the second sentence I wrote with your point? So ``That is, similar translations (e.g., the angle between chord-pair vectors) can be used to move between functional chords in different keys because the distance between chords (such as I-V) is preserved regardless of the key.''  Or is that too wordy? :) }  \dorien{Ah, that sounds great!} \ch{Looks good!}%Proximity (cosine distance) between chords in word2vec space reflects the fact that the model was able to learn about the functional role and relationships between chords (see Fig.~\ref{fig:chord_distance}) \ch{Remove the previous sentence about proximity since we only use angles.}. %Further, 
As we can see from the angle between chord-pair vectors across keys, the model successfully learns the relationship between functional chords in different keys: 
%\kat{[I know this result is from the previous section in the paper (before Analogy), but can't we say the following about analogy too? referring to results of Fig 5]:} 
%the \kat{distance between I-IV and I-V chords is smaller than for I-IIb chords across keys, as shown in Figure~\ref{fig:chord_distance} [UPDATE: if analogy is only about translating via angles, then I guess we shouldn't include the finding about cos distance]} \dorien{I think it's ok to bring two things together in the conclusion} \ch{The discussion on the distance between chords is already in the previous paragraph (I and V have smaller cosine distance than I and II chords).} \kat{Okay, I think this is just causing confusion, so I'm going to remove that part about distance and focus on angle for analogy results}, and 
the angle between I-V and i-v chord-pair vectors is more consistent than for I-vi chord-pair vectors across keys (see Figures~\ref{fig:vector_C_G} and ~\ref{fig:vector_others}). %\ch{Should it be "that the angle between I-V and i-v chord-pair vectors is more consistent than for I-vi chord-pair vectors across keys"?} \kat{Ah, you're right, we definitely shouldn't say smaller.. Yes, I think your text is better. I've updated the paragraph.}

Given the potential of word2vec to capture musical characteristics, we performed an initial exploration of how a semantic vector space model may be used for music generation. The approach we tested involved a new strategy for replacing musical slices with tonally-similar slices based on word2vec cosine similarity. While it is not within the scope of the current paper to explore the quality of the generated music, our approach describes a new way in which word2vec could be a useful tool for future automatic composition tools.

%Slices were replaced based on word2vec context similarity also presents close tonal distance compared to the original. 

%In the future, word2vec could be used to... 

%\kat{might need your inspiration on this, Ching-Hua :)  Maybe tie in generation, eg., for composers/musicians. Also could be used for (or as input to) models of similarity and structure, such as Dorien's tension model, or Elaine's model, etc... that is, a word2vec approach could be a new way of capturing \textit{semantic similarity} in music.} 

This paper demonstrates the promising capabilities of word2vec for modeling basic musical concepts. In the future, we plan to further explore word2vec's ability to model complex relationships by comparing it with mathematical models of tonality, such as the one described in~\citep{chew2014mathematical}. We also plan to combine word2vec with sequential modeling techniques (e.g., recurrent neural networks) to model structural concepts like musical tension~\citep{herremans2015generating}. The results of such studies could provide a new way of capturing \textit{semantic similarity} and \textit{compositional style} in music. We will also continue to explore the use of word2vec in music generation by further investigating aspects of music similarity and style. 
In future work, it would also be interesting to investigate how word2vec may be integrated as an automatic feature extraction model into existing models, such as RNNs and LSTMs. This could be done, for instance, by using the learned word2vec word embeddings as input for these models. 
The reader is invited to further build upon the model of this project, which is available online.\footnote{\url{https://sites.google.com/view/chinghuachuan/}}
%\ch{Feel free to edit/add more to this :)}  
%\kat{Thanks! Done revising.}

%Given the multidimensional nature of the vector space, it would be interesting to compare the music word2vec space to other higher-dimensional geometric models such as the Spiral Array~\citep{chew2014mathematical} based on music theory in future research. 
%\end{itemize} 

%\rewrite{an embedded model that combines both word2vec with, for instance, a long-short term memory recurrent neural network based on musical features, would offer a more complete way to model music. The TensorFlow code used in this research is available online\footnote{http://dorienherremans.com/word2vec}.}

%\dorien{too bad we can't cite our ismir yet here} 

%\dorien{I think we shouldn\t worry about paper limits, as in the end, we decide on the length... right?}

\begin{acknowledgements}
This research was partly supported through SUTD Grant no. SRG ISTD 2017 129. 
\end{acknowledgements}

\small 
\noindent
\textbf{Conflict of interest} The authors declare that they have no conflict of interest.

\bibliographystyle{spbasic}      % basic style, author-year citations
\bibliography{paper}   % name your BibTeX data base

\clearpage

\appendix

\FloatBarrier
\section{Appendix}

\begin{figure}[h] \centering
\includegraphics[width=\textwidth]{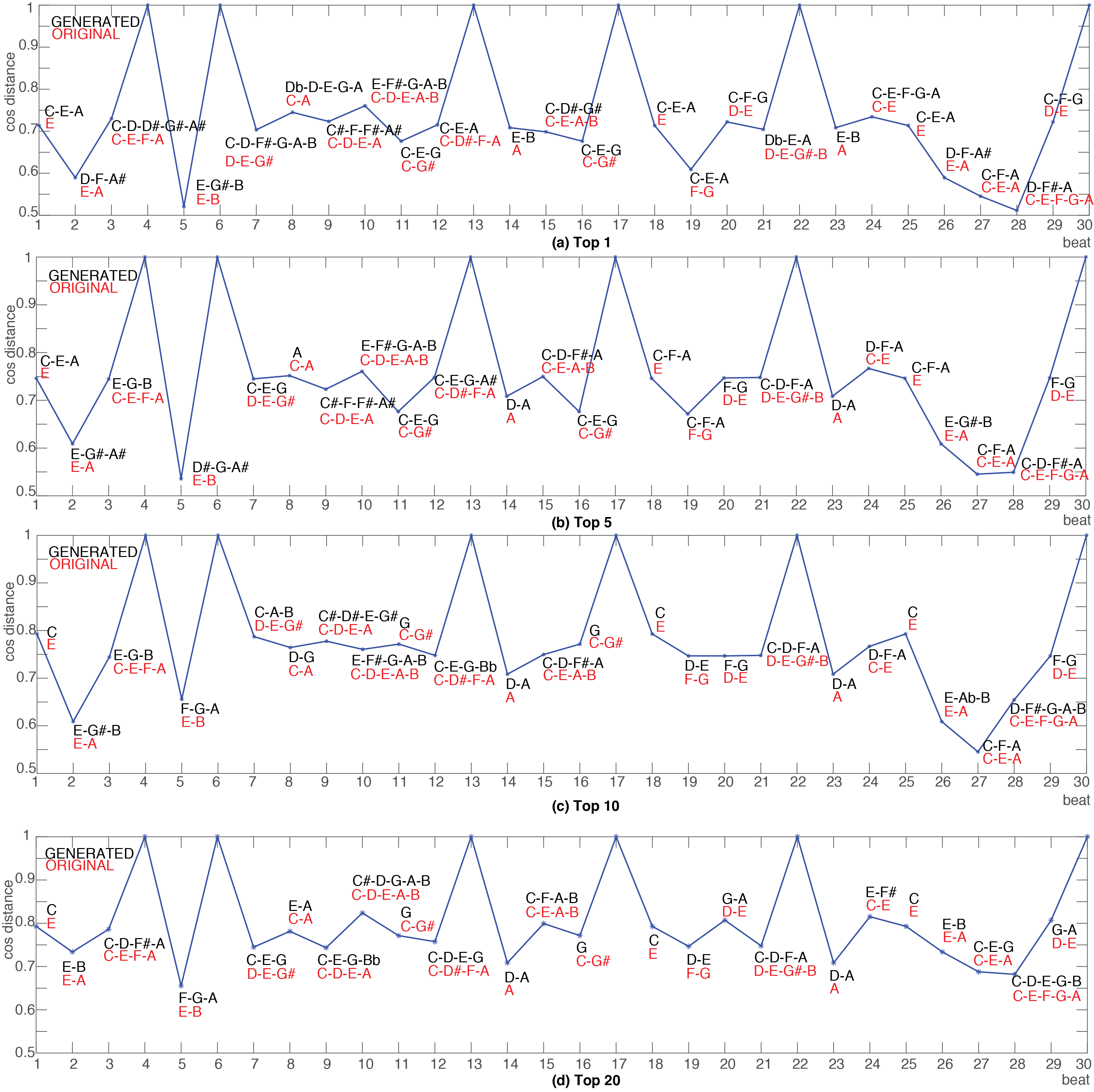}
\caption{Generated slices and their cosine distance to the original slices from Chopin's Mazurka Op. 67 No. 4, using (a) top 1, (b) top 5, (c) top 10, and (d) top 20 slices for the search in music word2vec space. Note that as the value of \textit{n} increases (e.g., moving from figure (a) down to (d)), the number of pitches outside of the key (see generated pitches in black) decreases.} %\ch{I am thinking about putting this figure in the appendix. It has too many details and I don't want to explain all in the paper to make it hard to read, but I also want to provide this for the reader who is interested to see more. What do you think?}
%\dorien{sounds good. We can even think of putting the algorithms in appendix if the layout doesn't work out well.} \dorien{The appendix was not defined in the latex template, so I did it in my own way, hopefully ok} \ch{It looks great! Thanks!}
\label{fig:chopin_details}
\end{figure}

% \begin{thebibliography}{}
% %
% % and use \bibitem to create references. Consult the Instructions
% % for authors for reference list style.
% % Format for Journal Reference
% Author, Article title, Journal, Volume, page numbers (year)
% Author, Book title, page numbers. Publisher, place (year)
% \end{thebibliography}

\end{document}

%% file: graph.tex
% \documentclass[varwidth]{standalone}[2011/12/21]
% \usepackage{pgfplots}

%  \pgfplotsset{compat=1.3} 
% %define the plots and nbh formatting
% \newcommand{\nbh}[1]{\textsf{#1}}

% \begin{document}

\def\layersep{2.5cm}

\begin{tikzpicture}[scale=1, transform shape,shorten >=1pt,-,draw=black!50, node distance=\layersep]
    \tikzstyle{every pin edge}=[<-,shorten <=1pt]
    \tikzstyle{neuron}=[rectangle,fill=black!25,minimum size=17pt, minimum width=24pt, inner sep=0pt]
        \tikzstyle{longneuron}=[rectangle,fill=black!25, minimum width=17pt,minimum height=85pt,inner sep=0pt]
    \tikzstyle{input neuron}=[neuron, fill=green!50];
    \tikzstyle{output neuron}=[neuron, fill=purple!50];
    \tikzstyle{output neuron invis}=[neuron, fill=white];
    \tikzstyle{hidden neuron}=[longneuron, fill=blue!50];
    \tikzstyle{annot} = [text width=4em, text centered]

    % Draw the input layer nodes
%    \foreach \name / \y in {0}
%    % This is the same as writing \foreach \name / \y in {1/1,2/2,3/3,4/4}
%        \node[input neuron] (I-\name) at (0,-\y) {};

    % Draw the hidden layer nodes
    
%        \node[hidden neuron,right of=I-0, node distance=2.5cm] (H-3) {};
%     \foreach \name / \y in {1,...,5}
%         \path[yshift=0.5cm]
%             node[hidden neuron] (H-\name) at (\layersep,-\y cm) {};

    % Draw the output layer node
%\node[anchor=south west] (myfirstpic) at (0,0)    

      \node[output neuron] (O) at (0,0) {};
%\node[anchor=south west] (output neuron) at (0,0) (O) {}:
  %  \node[output neuron,right of=O, node distance=2cm] (O1) {};  
\node[output neuron, right of=O, node distance=1.4cm] (O2) {};           
\node[output neuron,right of=O2, node distance=1.4cm] (O3) {};                     
 \node[output neuron,right of=O3, node distance=1.4cm] (O4) {};
\node[output neuron,right of=O4, node distance=1.4cm] (O5) {};                     
 \node[output neuron,right of=O5, node distance=1.4cm] (O6) {};
\node[output neuron,right of=O6, node distance=1.4cm] (O7) {};                     
 %\node[output neuron,right of=O7, node distance=1.4cm] (O8) {};

%         \foreach \name / \y in {1,...,4}
%     % This is the same as writing \foreach \name / \y in {1/1,2/2,3/3,4/4}
%         \node[output neuron, pin={[pin edge={->}]right:Output}, right of=H-3] (O) {};

    % Connect every node in the input layer with every node in the
    % hidden layer.
    
     \draw [thick, black,decorate,decoration={brace,amplitude=10pt,mirror},xshift=0.4pt,yshift=-0.4pt](.8,-.6) -- (3.4,-.6) node[black,midway,yshift=-0.6cm] {};

     \draw [thick, black,decorate,decoration={brace,amplitude=10pt,mirror},xshift=0.4pt,yshift=-0.4pt](5.1,-.6) -- (7.6,-.6) node[black,midway,yshift=-0.6cm] {};

%    \path (H-3) edge (O2);
%    \path (H-3) edge (O3);
%    \path (H-3) edge (O4);
%    \path (H-3) edge (O1);
%    \path (I-0) edge (H-3);
%     \foreach \source in {1,...,4}
%         \foreach \dest in {1,...,5}
%             \path (I-\source) edge (H-\dest);

    % Connect every node in the hidden layer with the output layer
%     \foreach \source in {1,...,5}
%         \path (H-\source) edge (O);

    % Annotate the layers
%    \node[annot,above of=H-3, node distance=3cm] (hl) {Projection};
%        \node[annot,below of=H-3, node distance=2cm] (h5) {$n$-dim.};
%    \node[annot,top of=hl] {04} {test};
%    \node[annot,right of=hl] {Output};
    \node[annot,above of=O4, node distance=.7cm] {$w_{t}$};
    \node[annot,below of=O3, node distance=1.2cm] {$w_{t-1}$};
    \node[annot,below of=O2, node distance=1.2cm] {$w_{t-2}$};
    \node[annot,below of=O5, node distance=1.2cm] {$w_{t+1}$};
    \node[annot,below of=O6, node distance=1.2cm] {$w_{t+2}$};
%    \node[annot,right of=O1, node distance=1cm] {$w_{t+1}$};
%    \node[annot,right of=O2, node distance=1cm] {$w_{t+2}$};
%    \node[annot,right of=O3, node distance=1cm] {$w_{t-1}$};
%    \node[annot,right of=O4, node distance=1cm] {$w_{t-2}$};
\end{tikzpicture}